%% file: main.tex
\documentclass[10pt,journal,compsoc]{IEEEtran}

\IEEEoverridecommandlockouts
\usepackage{amsmath,amssymb,amsfonts}
\usepackage{algcompatible}
\usepackage[ruled,linesnumbered]{algorithm2e}
\usepackage{color}
\usepackage{graphicx}
\usepackage{textcomp}
\usepackage[dvipsnames]{xcolor}
\usepackage{paralist}
\usepackage{subcaption}
\usepackage{multirow}
\usepackage{graphics}
\usepackage{url}
\usepackage{cancel}
\usepackage{makecell}
\def\BibTeX{{\rm B\kern-.05em{\sc i\kern-.025em b}\kern-.08em
    T\kern-.1667em\lower.7ex\hbox{E}\kern-.125emX}}
\usepackage[colorinlistoftodos,prependcaption,textsize=footnotesize]{todonotes}

\usepackage{mathtools}
\DeclarePairedDelimiter{\ceil}{\lceil}{\rceil}
\DeclarePairedDelimiter{\floor}{\lfloor}{\rfloor}

\usepackage[noadjust]{cite}

\let\oldnl\nl% Store \nl in \oldnl
\newcommand{\nonl}{\renewcommand{\nl}{\let\nl\oldnl}}% Remove line number for one line

\usepackage{amsmath,bm}

\usepackage{xcolor,cancel}

\newcommand{\gpk}{\texttt{GPK}} 
\newcommand{\lpk}{\texttt{LPK}}
\newcommand{\ipk}{\texttt{IPK}} 

\begin{document}
% This command plus prefix in bib file gives "X et al" for refs.
\bstctlcite{IEEEexample:BSTcontrol}

%% Macro to produce version without comments
\newif\iffinal
\finalfalse

%% Macros for comments and todos
\iffinal
  \newcommand\guidance[1]{}
  \newcommand\writer[1]{}
  \newcommand\kshitij[1]{}
  \newcommand\scott[1]{}
  \newcommand\matthew[1]{}
  \newcommand\jong[1]{}
  \newcommand\ian[1]{}
  \newcommand\todd[1]{}
  \newcommand\igor[1]{}
  \newcommand\dave[1]{}
\else
  \definecolor{amber}{rgb}{1.0, 0.49, 0.0}
  \definecolor{darkblue}{rgb}{0,0,0.5}
  \definecolor{crimson}{rgb}{0.83, 0.0, 0.25}
  \definecolor{darkgreen}{rgb}{0,0.5,0}
  \definecolor{purple}{rgb}{0.5,0,0.5}
  \definecolor{brown}{rgb}{0.65, 0.16, 0.16}
  \definecolor{caribbeangreen}{rgb}{0.0, 0.8, 0.6}
  \definecolor{carmine}{rgb}{0.59, 0.0, 0.09}
  \definecolor{champagne}{rgb}{0.97, 0.91, 0.81}
  \definecolor{classicrose}{rgb}{0.98, 0.8, 0.91}
  \definecolor{corn}{rgb}{0.98, 0.93, 0.36}
  \definecolor{cornflowerblue}{rgb}{0.39, 0.58, 0.93}
  \definecolor{darkelectricblue}{rgb}{0.33, 0.41, 0.47}
  \newcommand\guidance[1]{\todo[inline]{#1}}
  \newcommand\writer[1]{\color{red} -- #1}
  \newcommand\kshitij[1]{{\todo[inline,color=green]{Kshitij: #1}}}
  \newcommand\scott[1]{\todo[inline,color=red]{Scott: #1}}
  \newcommand\matthew[1]{{\todo[inline,color=champagne]{Matthew: #1}}}
  \newcommand\jong[1]{{\todo[inline,color=amber]{Jong: #1}}}
  \newcommand\ian[1]{{\todo[inline,color=corn]{Ian: #1}}}
  \newcommand\todd[1]{{\todo[inline,color=classicrose]{Todd: #1}}}
  \newcommand\igor[1]{{\todo[inline,color=caribbeangreen]{Igor: #1}}}
  \newcommand\dave[1]{{\todo[inline,color=orange]{Dave: #1}}}
  \newcommand\jieyang[1]{{\todo[inline,color=caribbeangreen]{Jieyang: #1}}}
\fi

\title{Scalable Multigrid-based Hierarchical Scientific Data Refactoring on GPUs}

\author{Jieyang Chen,
Lipeng Wan, 
Xin Liang,
Ben Whitney,
Qing Liu,
Qian Gong,
David Pugmire,
Nicholas Thompson,
Jong Youl Choi,
Matthew Wolf,
Todd Munson,
Ian Foster,
Scott Klasky
%~\IEEEmembership{Senior Member,~IEEE}

\IEEEcompsocitemizethanks{\IEEEcompsocthanksitem J. Chen, L. Wan, B. Whitney, Q. Gong, D. Pugmire, N. Thompson, J. Choi, M. Wolf, S. Klasky are with the Computer Science and Mathematics Division, Oak Ridge National Laboratory, Oak Ridge, TN 37830.\protect\\
E-mail: \{chenj3, wanl, whitneybe, gongq, pugmire, thompsonna, choij, wolfmd, klasky\}@ornl.gov
\IEEEcompsocthanksitem X. Liang is with the Missouri University of Science and Technology, Rolla, MO 65409.\protect\\
E-mail: \{xliang\}@mst.edu
\IEEEcompsocthanksitem T. Munson, I. Foster are with the Argonne National Laboratory, Lemont, IL 60439.\protect\\
E-mail: \{tmunson, foster\}@mcs.anl.gov
}% <-this % stops an unwanted space
%\thanks{
% Manuscript received Nov. 10, 2020. 
%\textit{}
%}

% \\
% \IEEEauthorblockA{
% Oak Ridge National Laboratory, Oak Ridge, TN, USA \\
% }
% \IEEEauthorblockA{\IEEEauthorrefmark{1}
% Missouri University of Science and Technology, Rolla, MO, USA \\
% }
% \IEEEauthorblockA{\IEEEauthorrefmark{2}
% New Jersey Institute of Technology, Newark, NJ, USA \\
% }
% \IEEEauthorblockA{\IEEEauthorrefmark{3}
% Argonne National Laboratory, Lemont, IL, USA \\
% }

% \IEEEauthorblockA{\IEEEauthorrefmark{4}
% University of Chicago, Chicago, IL, USA \\
% }

% \{chenj3, wanl, whitneybe, pugmire, thompsonna, choij, wolfmd, klasky\}@ornl.gov\\
% xlang@mst.edu qliu@njit.edu 
% \{tmunson, foster\}@anl.gov
}

\markboth{IEEE Transactions on Parallel and Distributed Systems}{Scalable Miultigrid-based Data Refactoring}

\IEEEtitleabstractindextext{%
\input{tex/abstract.tex}
}

\maketitle
\IEEEdisplaynontitleabstractindextext
\IEEEpeerreviewmaketitle
\linespread{0.95}
\selectfont

%150-word short abstract

% \input{tex/abstract.tex}
% \begin{IEEEkeywords}
% Multigrid, Data refactoring, GPU
% \end{IEEEkeywords}

\setlength{\textfloatsep}{0pt}

%Ref this paper?~\cite{alexander2020exascale}

\input{tex/introduction.tex}
\input{tex/backgrounds.tex}

\input{tex/design.tex}
\input{tex/evaluation.tex}

\input{tex/showcase}

\input{tex/related_works.tex}
\input{tex/conclusion.tex}
%\paragraph*{Acknowledgment}
\ifCLASSOPTIONcompsoc
  \section*{Acknowledgments}
\else
  \section*{Acknowledgment}
\fi

This work was made possible by support from the Department of Energy's Office of Advanced Scientific Computing Research, including via the CODAR and ADIOS Exascale Computing Project (ECP) projects.  This research used resources of the Oak Ridge Leadership Computing Facility, a DOE Office of Science User Facility supported under Contract DE-AC05-00OR22725.

\bibliographystyle{IEEEtran}
\bibliography{ref}
\vskip -2\baselineskip plus -1fil
\begin{IEEEbiography}[{\vspace{-11mm}\includegraphics[width=0.75in,height=1in,clip,keepaspectratio]{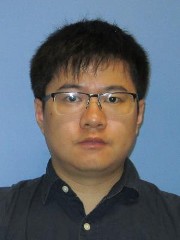}}]{Jieyang Chen} is a Computer Scientist at Computer Science and Mathematics Division at Oak Ridge National Laboratory.
He received his master and Ph.D. degrees in Computer Science from University of California, Riverside in 2014 and 2019. His research interests include high-performance computing, parallel and distributed systems, and big data analytics.
\end{IEEEbiography}
\vspace{-18mm}
%\vskip -2\baselineskip plus -1fil
\begin{IEEEbiography}[{\vspace{-11mm}\includegraphics[width=0.75in,height=1in,clip,keepaspectratio]{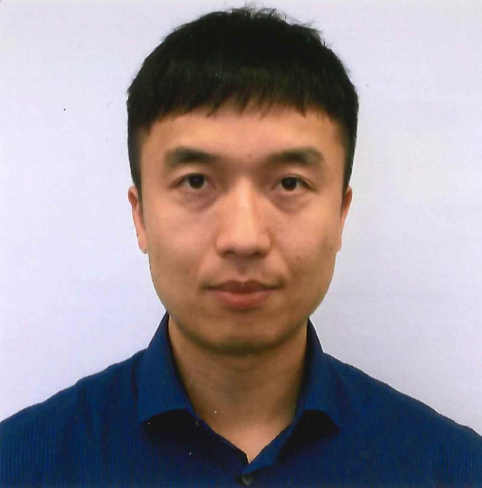}}]{Lipeng Wan} is a Computer Scientist in the Computer Science and Mathematics Division at Oak Ridge National Laboratory. He received his Ph.D. degree in computer science from the University of Tennessee, Knoxville in 2016. His research mainly focuses on scientific data management and high-performance computing.
\end{IEEEbiography} 
\vspace{-18mm}
%\vskip -2\baselineskip plus -1fil
\begin{IEEEbiography}[{\vspace{-11mm}\includegraphics[width=0.75in,height=1in,clip,keepaspectratio]{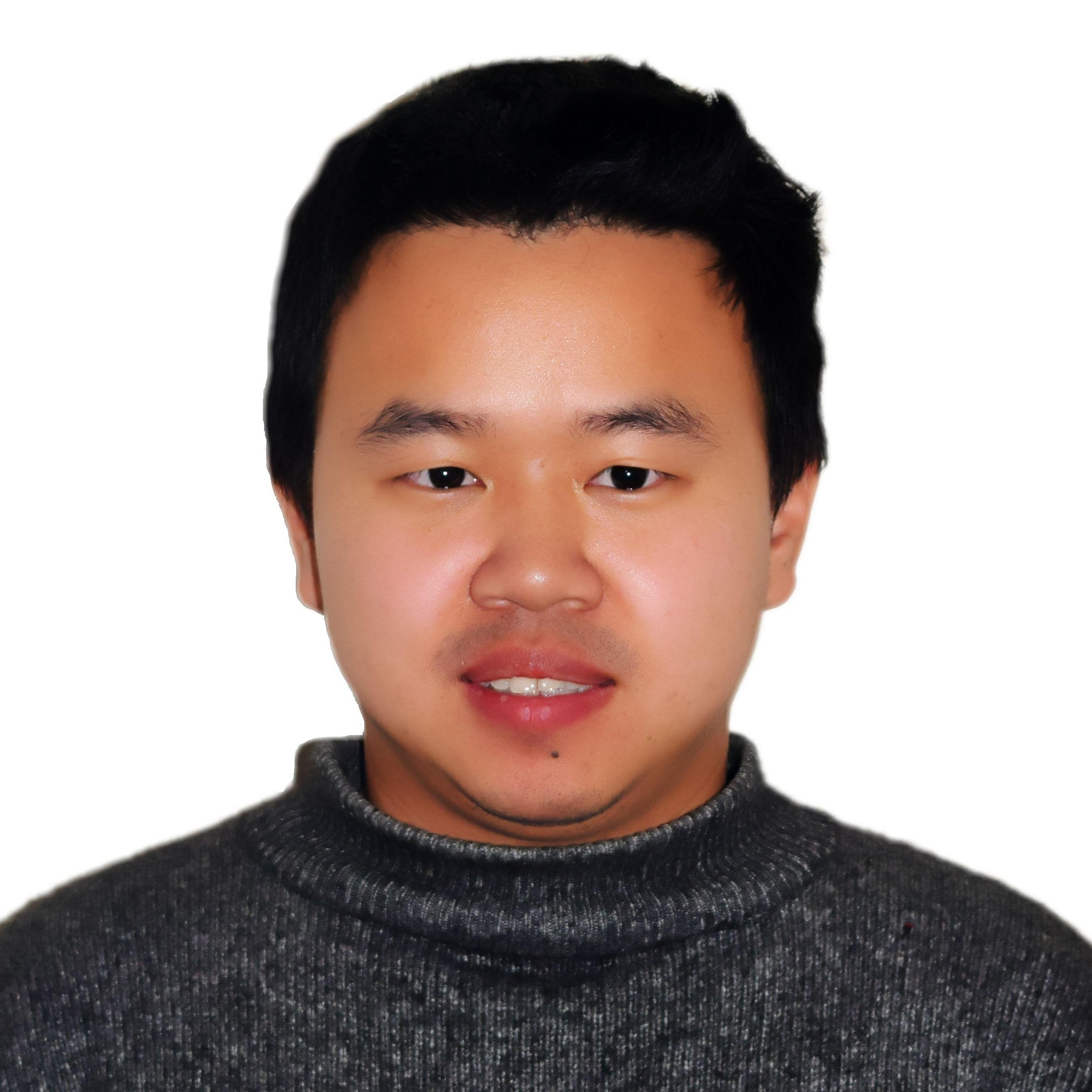}}]{Xin Liang} is an assistant professor with the Department of Computer Science at Missouri University of Science and Technology. He received his Ph.D. degree from University of California, Riverside in 2019.
%and his bachelor's degree from Peking University in 2014. 
His research interests include high-performance and distributed computing, 
%parallel and distributed systems,
data management and reduction, and cloud computing .
%big data analytic, scientific visualization, and cloud computing. 
\end{IEEEbiography}
\vspace{-18mm}
\begin{IEEEbiography}[{\vspace{-11mm}\includegraphics[width=0.75in,height=1in,clip,keepaspectratio]{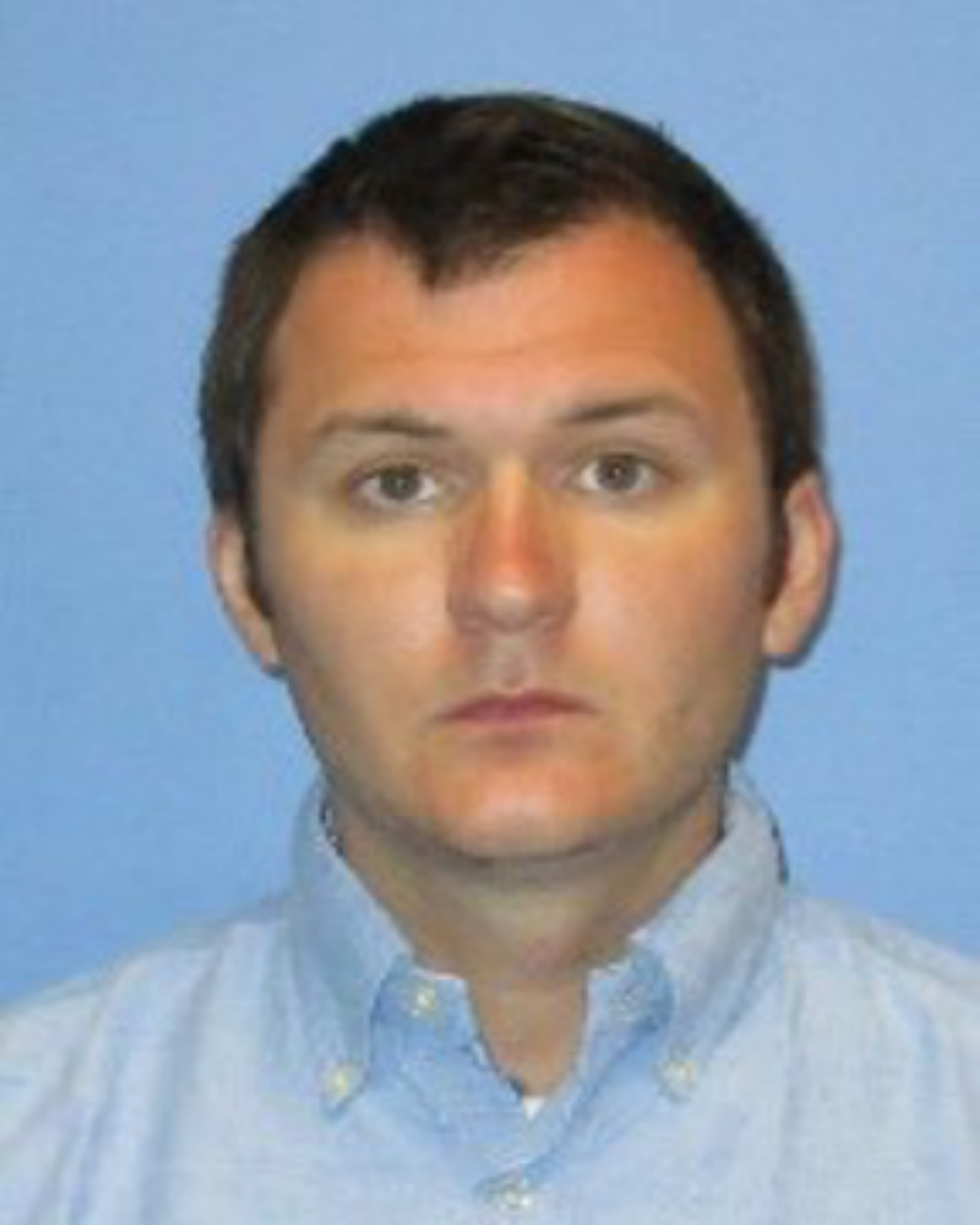}}]{Ben Whitney} is a postdoctoral research associate in the Computer Science and Mathematics Division at Oak Ridge National Laboratory.
He received his Ph.D. degree in applied mathematics from Brown University in 2018.
His research interests include data compression, numerical methods, and scientific software development.
\end{IEEEbiography}
\vspace{-18mm}
\begin{IEEEbiography}[{\vspace{-11mm}\includegraphics[width=0.75in,height=1in,clip,keepaspectratio]{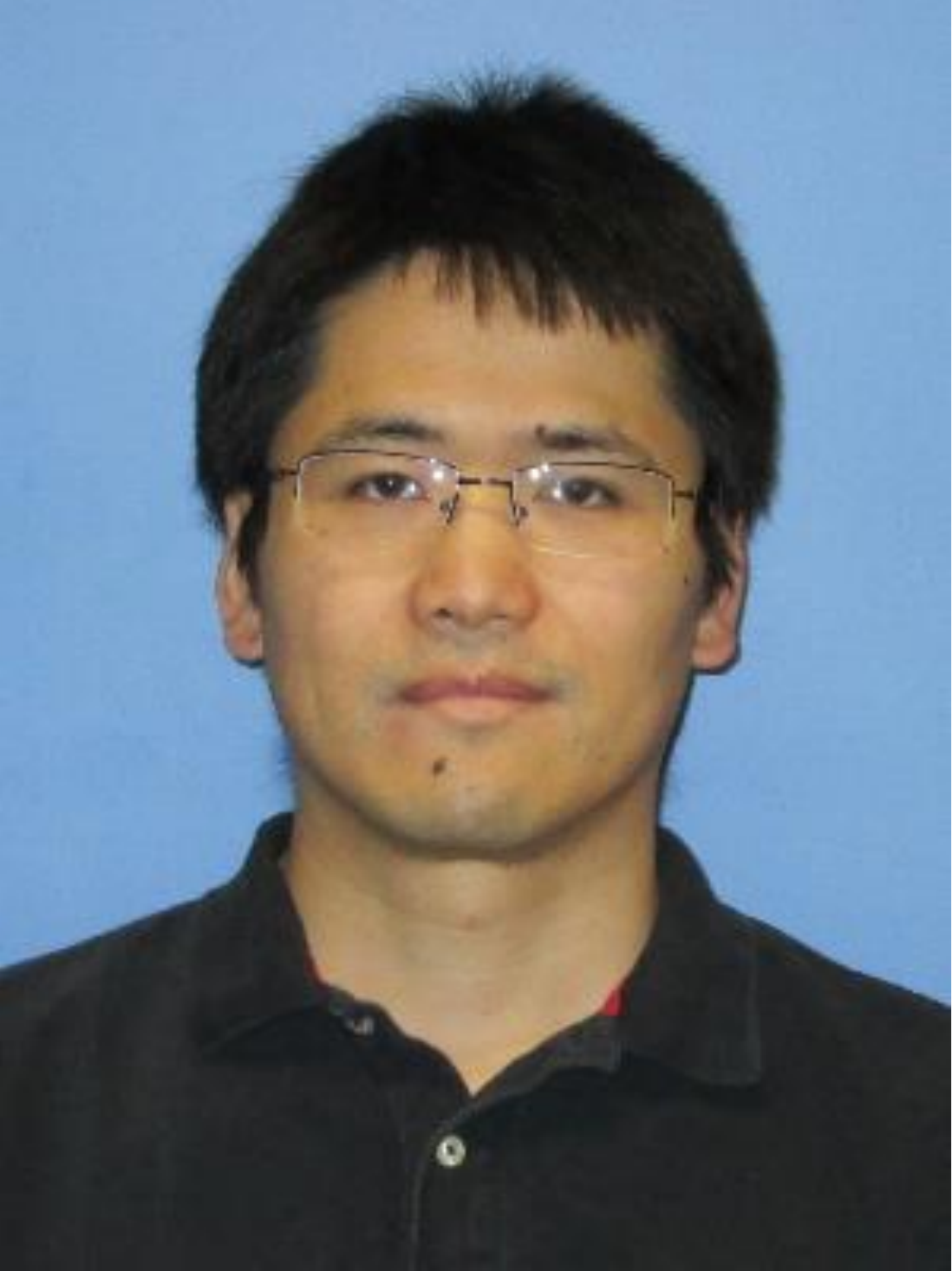}}]{Qing Liu} is an Assistant Professor in the Department of Electrical and Computer Engineering at NJIT. He has joint faculty appointment with Oak Ridge National Laboratory. 
%Prior to that, he was a staff scientist at Science Data Group, Oak Ridge National Laboratory. 
He received his Ph.D. in Computer Engineering from the University of New Mexico in 2008. His areas of interest include high-performance computing, data science, and networking. 
%In 2013 He won an R\&D 100 award for the development of the Adaptable I/O System for Big Data.
\end{IEEEbiography}
\vspace{-18mm}
\begin{IEEEbiography}[{\vspace{-11mm}\includegraphics[width=0.75in,height=1in,clip,keepaspectratio]{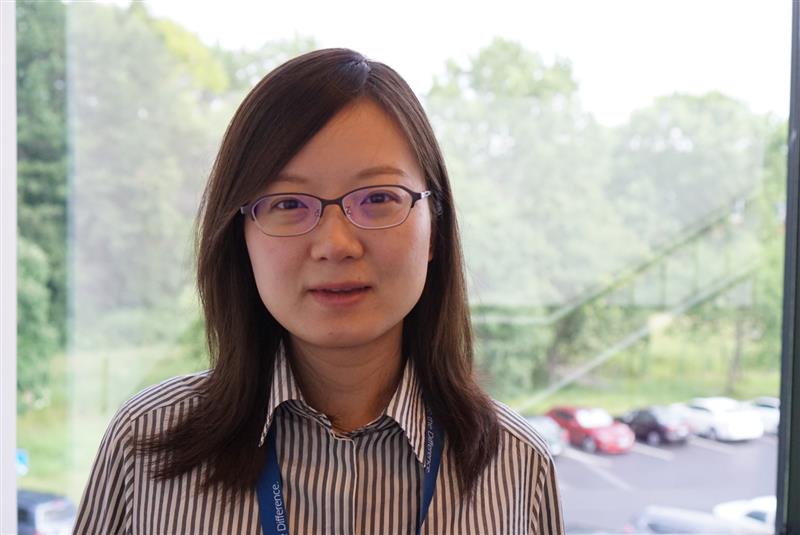}}]{Qian Gong} is a computer scientist at Oak Ridge National Laboratory. Before that, she was a researcher at Fermi National Accelerator Laboratory. She received her Ph.D in Electrical and Computer Engineering from Duke University in 2017. Her current research endeavors involve data reduction, and parallel computing.
\end{IEEEbiography}
\vspace{-18mm}
\begin{IEEEbiography}[{\vspace{-11mm}\includegraphics[width=0.75in,height=1in,clip,keepaspectratio]{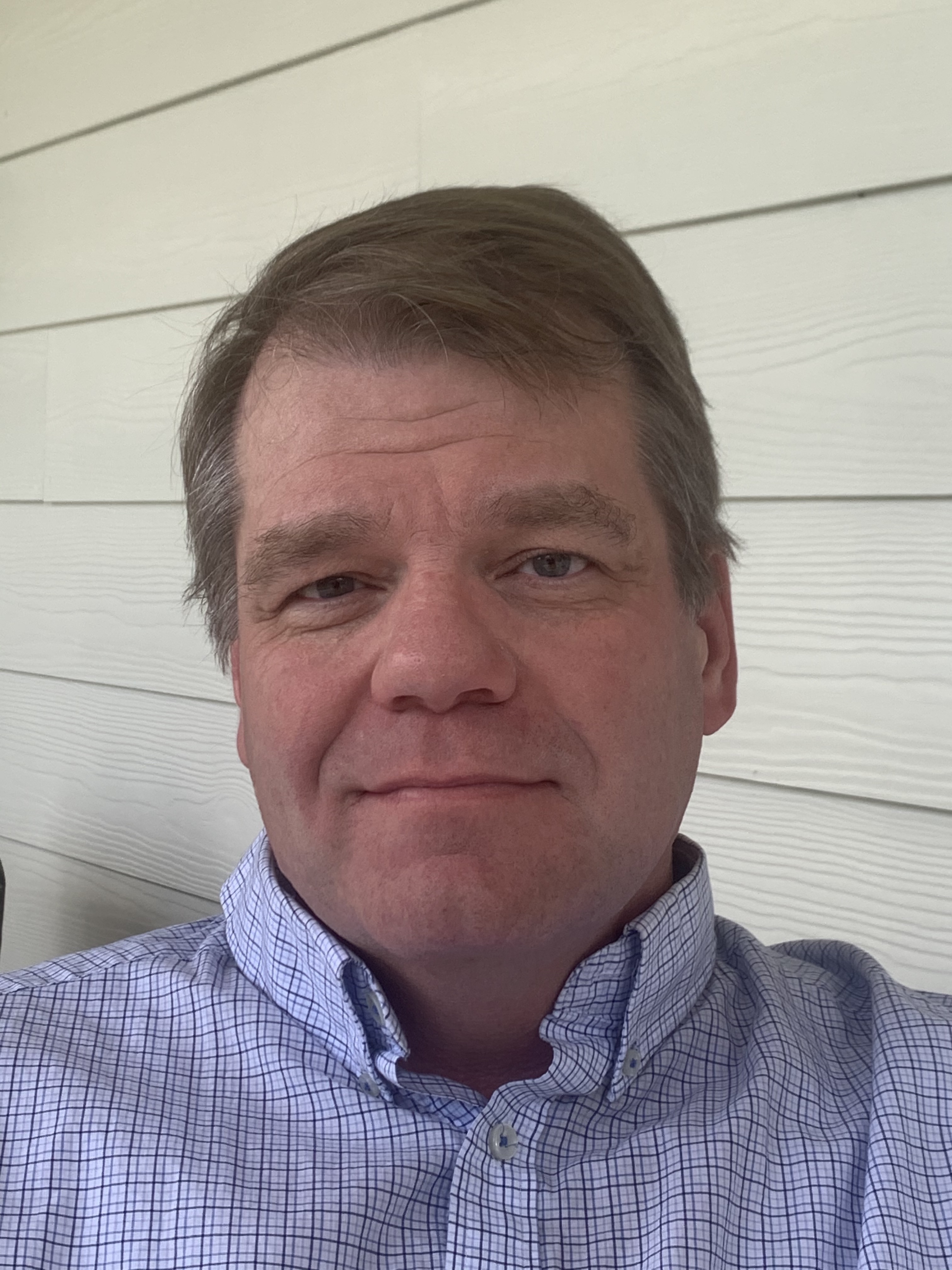}}]{David Pugmire} is a Senior Research Scientist and Visualization Group Leader in the Data and AI Section at Oak Ridge National Laboratory and Joint Faculty Professor at the University of Tennessee. He received his Ph.D. from the University of Utah in 2000. 
%Before joining ORNL, he was a Research Scientist at Los Alamos National Laboratory. 
His research interests include scalable visualization on high performance computing systems. 
%In 2006 he won an R\&D 100 award for an NPU-based image compositor. He has published numerous papers in high-quality conferences and journals, including SC, IEEE Visualization, EuroVis, IPDPS, BigData and TVCG. He is a member of the ACM.
\end{IEEEbiography}
\vspace{-18mm}
\begin{IEEEbiography}[{\vspace{-11mm}\includegraphics[width=0.75in,height=1in,clip,keepaspectratio]{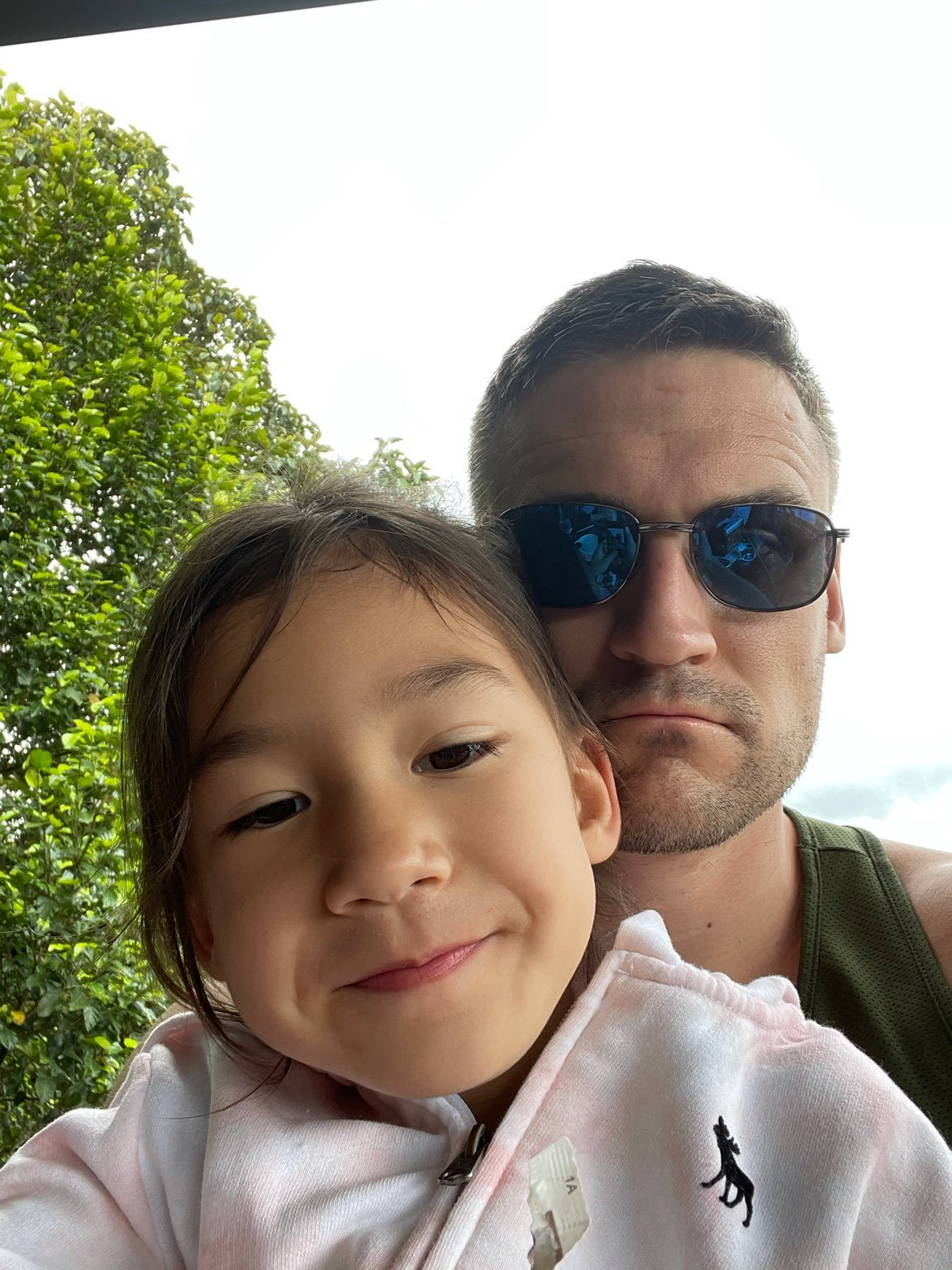}}]{Nicholas Thompson} is a numerical software developer at Oak Ridge National Lab, working on VTK-m, ADIOS, and Boost. His research interests include solving numerical problems, scientific visualization and data management. 
\end{IEEEbiography}
\vspace{-18mm}
\begin{IEEEbiography}[{\vspace{-11mm}\includegraphics[width=0.75in,height=1in,clip,keepaspectratio]{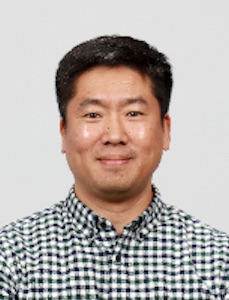}}]{Jong Youl Choi} is a Computer  Scientist in the Computer Science and Mathematics Division at Oak Ridge National Laboratory. He earned his Ph.D. degree in Computer Science at Indiana University Bloomington. His areas of research span data mining and machine learning algorithms, high-performance computing, and parallel and distributed systems.
%, focusing on researching and developing data-centric machine learning algorithms, in situ/in-transit data processing methods, and data management for coupled simulation.
\end{IEEEbiography}
\vspace{-18mm}
\begin{IEEEbiography}[{\vspace{-11mm}\includegraphics[width=0.75in,height=1in,clip,keepaspectratio]{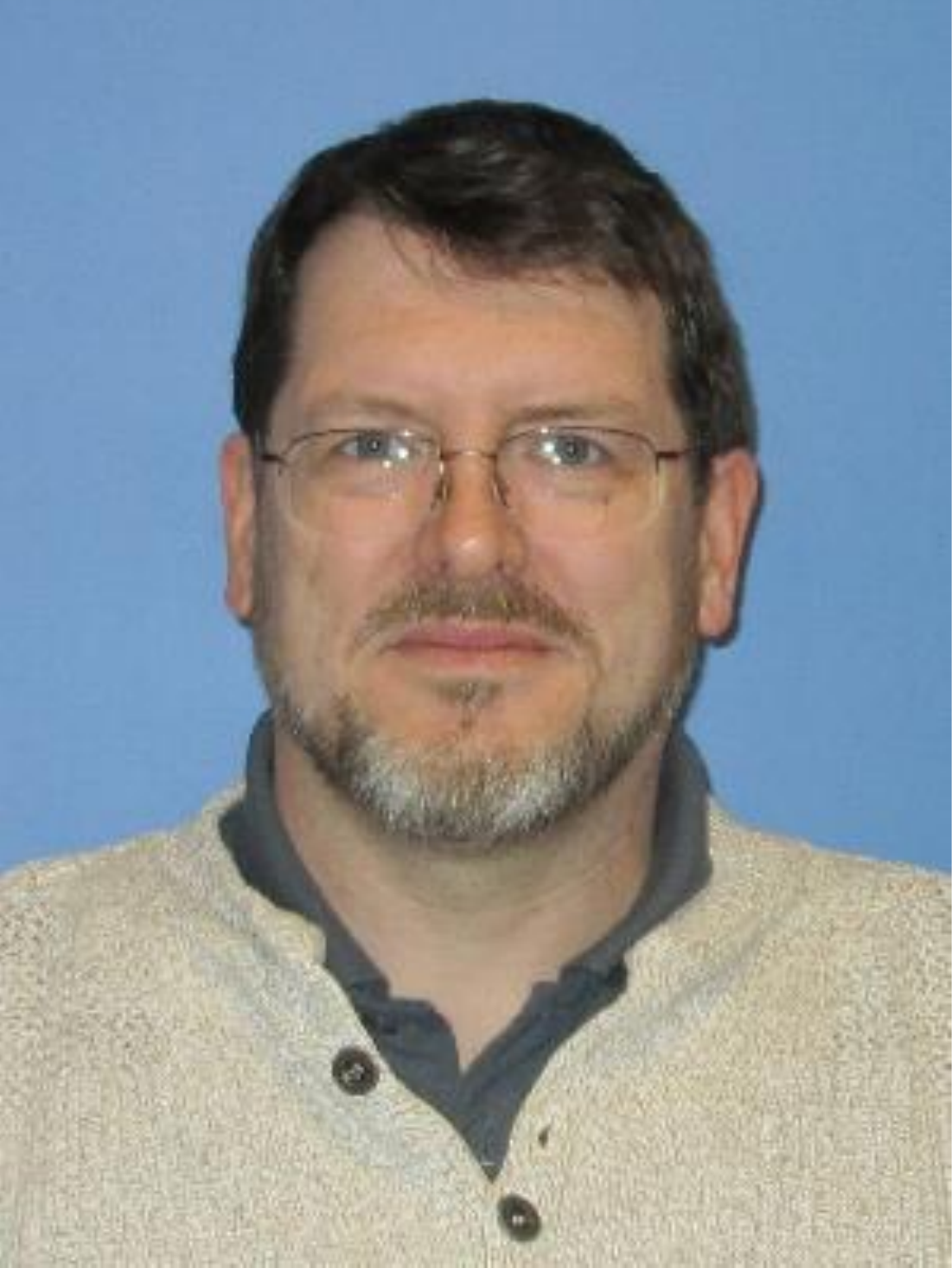}}]{Matthew Wolf} is a Senior Computer Scientist in the Computer Science and Mathematics Division at Oak Ridge National Laboratory.  
%Before joining ORNL full-time, he was a senior research scientist at Georgia Institute of Technology with a joint appointment to ORNL.  
His research interests include high performance computing, adaptive I/O and messaging middleware, and in situ analysis and visualization systems for science.  
%He has advised and co-advised numerous students who have gone on to careers in industry, academia, and the national laboratories, as well as publishing papers in prominent conferences and journals. 
\end{IEEEbiography}
\vspace{-18mm}
\begin{IEEEbiography}[{\vspace{-11mm}\includegraphics[trim=25 120 40 15,width=0.75in,height=1in,clip,keepaspectratio]{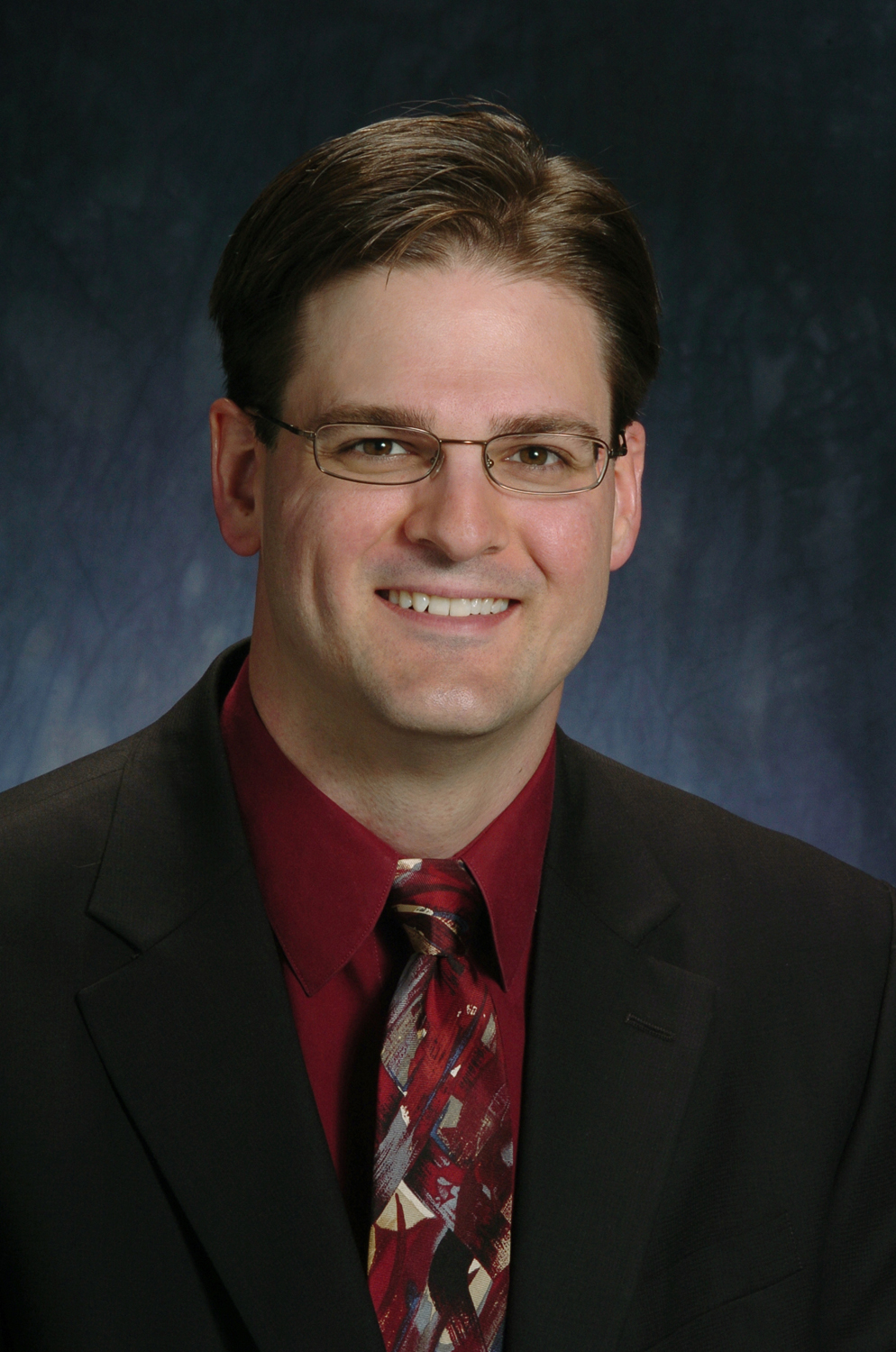}}]{Todd Munson} is a senior computational scientist
at Argonne National Laboratory
% , a senior scientist
% at the Consortium for Advanced Science and 
% Engineering at the University of Chicago, 
and the Software Ecosystem and Delivery Control
Account Manager for the U.S.\ DOE Exascale Computing Project.
He received his Ph.D from the University of Wisconsin at 
Madison in 2000. His interests range from numerical 
methods
%for nonlinear optimization and variational 
%inequalities 
to workflow optimization for online 
data analysis and reduction.
\end{IEEEbiography}
\vspace{-18mm}
\begin{IEEEbiography}[{\vspace{-11mm}\includegraphics[width=0.75in,height=1in,clip,keepaspectratio]{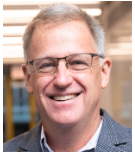}}]{Ian Foster} is a senior Scientist and Distinguished Fellow, and also director of the Data Science and Learning Division, at Argonne National Laboratory, and the Arthur Holly Compton Distinguished Service Professor at the University of Chicago. His research deals with distributed, parallel, and data-intensive computing.
\end{IEEEbiography}
\vspace{-18mm}
\begin{IEEEbiography}[{\vspace{-11mm}\includegraphics[width=0.75in,height=1in,clip,keepaspectratio]{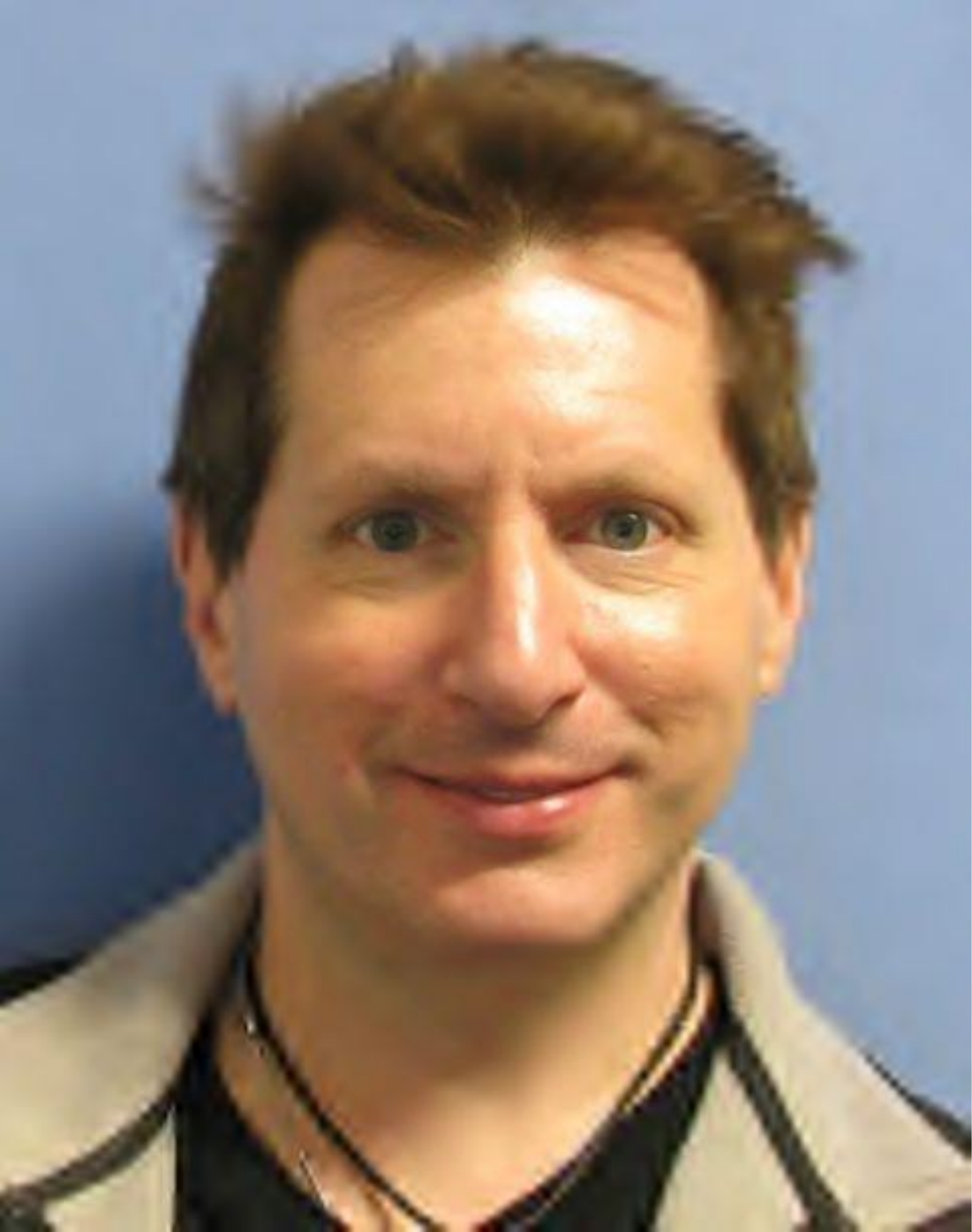}}]{Scott Klasky} is a distinguished scientist and Group Leader in the CSM Division at ORNL. 
He also
has a joint faculty appointment at the University of Tennessee, Knoxville and Georgia Tech.  
He received his Ph.D. in from the University
of Texas at Austin. %He is a senior Member of IEEE, and won an R\& D 100
%award for being the leader for the Adaptable I/O System (ADIOS). 
He has expertise in HPC, data management, workflow automation, data reduction, visualization and physics.
\end{IEEEbiography}

\end{document}

%% file: tex/abstract.tex
\begin{abstract}
Rapid growth in scientific data and a widening gap between computational speed and I/O bandwidth makes it increasingly infeasible to store and share all data produced by scientific simulations.
Instead, we need methods for reducing data volumes: ideally, 
methods that can scale data volumes adaptively so as to enable negotiation of performance and fidelity tradeoffs in different situations.
Multigrid-based hierarchical data representations hold promise as a solution to this problem, 
allowing for flexible conversion between different fidelities
so that, for example, data can be created at high fidelity and then transferred or
stored at lower fidelity via logically simple and mathematically sound operations. 
However, the effective use of such representations has been hindered until now by the relatively high costs of creating, accessing, reducing, and otherwise operating on
such representations.
We describe here highly optimized data refactoring kernels for GPU accelerators that enable efficient creation and manipulation of data in multigrid-based hierarchical forms.
We demonstrate that our optimized design can achieve up to 264 TB/s aggregated data refactoring throughput---92\% of theoretical peak---on 1024 nodes of the Summit supercomputer.
% We also specially optimized GPU data refactoring for consumer class GPU commonly used in edge systems, dense multi-GPU architecture used in HPC systems, and spatiotemporal data refactoring.
We showcase our optimized design by applying it to a large-scale scientific visualization workflow and the MGARD lossy compression software.

\end{abstract}

%% file: tex/introduction.tex
\IEEEraisesectionheading{\section{Introduction}}
%{ People to help \writer{Lipeng, Matthew, Dave, Copy-editing: Kevin or Todd (done)}}

%\ian{Cite CODAR if we can~\cite{CODAR2020}.}

\IEEEPARstart{W}{ith}
the dawn of the big data era, managing the massive volume of data generated by data-intensive applications becomes extremely challenging, particularly for scientific simulations~\cite{alexander2020exascale, Wan:2019} running on leadership-class high-performance computing (HPC) systems and experiments running on federated instruments and sensor platforms. 
For instance, the XGC dynamic fusion simulation \cite{ku2009full,chang2004numerical} from  the Department of Energy (DoE)'s Princeton Plasma Physics Laboratory can generate 1 PB per day when running on DoE's fastest supercomputers, and may soon generate 10 PB  per day.  
The Square Kilometer Array \cite{taylor2004science} plans to generate data at 1~PB/s within 10--20 years. Few storage systems can keep up with such data rates.
Moreover, even if all data could be stored, the high costs of 
processing them with standard multi-pass analysis routines often lead to significant degradation in overall scientific productivity.% \cite{Wan:2017,Wan:hpcc:2017}.

Current solutions for managing this overwhelming amount of complex and heterogeneous scientific data are usually passive and based on rules of thumb. 
For instance, scientists may decimate in time by reducing output data frequency by some arbitrary factor (e.g., writing data every 1000 simulation steps). 
Although such approaches can effectively reduce the amount of data written to storage, they increase the risk of missing novel scientific discoveries, as discarded data may contain important features. 
Moreover, the limited capacity of fast storage such as parallel file systems means that data are eventually moved to slower storage, such as archival storage systems.
For example, on Oak Ridge National Laboratory (ORNL)'s Summit supercomputer, data can only be kept on the parallel file system for 90 days before it is either moved to archival storage systems such as HPSS \cite{hpss} or permanently deleted. 
Once moved to archival storage, it can take weeks or longer to retrieve for analysis. 

There are no universal solutions to the many technical and domain-specific challenges of large data. In fact, from the domain scientist's perspective, being able to store large amounts of data may not lead to more scientific discoveries. The most valuable scientific insights often come from just a small portion of the original data.
%\ian{I wonder if this text confounds two ideas: that different parts of data may be of varying use (e.g., data in the tropics for a climate model, vs. different levels of accuracy (e.g., single precision is enough to solve a problem?}
%\jieyang{Yes. But the current algorithm can only address the first. The second is an ongoing work.}
%while the remaining data are less useful. 
Domain scientists would like to be able to retrieve only enough data to guarantee a certain accuracy, so that desired results can be produced when applying a type of analysis on the reduced data representation.
However, it is challenging to use existing data compression techniques, given differing data accuracy needs of different analyses, since data need to be compressed and stored separately for different needed accuracies, which brings high computational and storage costs.

%\jieyang{The following paragraph is what I rewrote for introducing the term 'data refactoring' with associated terms: 'decomposition', 'recomposition', and 'coefficient classes'. Does it read better?}
\emph{Data refactoring} is the capability of building a data representation in a hierarchical form such that a reader can easily, efficiently, and transparently access data at varying degrees of fidelity.
%reduce and reorganize the data intelligently based on their intentions so that: 1) they do not lose valuable data containing important features; and 2) they can quickly access needed data when they run their analysis routines.
%\ian{The requirements seem a bit vague to me. ``not losing any valuable data'' and ``accessing needed data fast'' are aspirations, not requirements, in the absence of a definition of valuable and needed. Also, maybe write, ''retain valuable data''}
%We call such capability \emph{intelligent scientific data refactoring}: a transformation of the internal representation of data such that data represented in an importance-first and progressive fashion with accuracy guarantee.
%Namely, more data are retrieved means they can represents the original data more accurately and vice versa.
%\ian{I think that a definition of data refactoring is needed. As far as I can tell, this is a term that you invented (refactoring usually means code refactoring).I am not sure yet what it means.}
%\matthew{I would suggest adding something like the following as the definition that Ian was asking for: "We use the term data refactoring in much the same way it is used for code or database refactoring: a transformation of the internal representation while preserving the ability to access and query data as before.  Here, we allow the refactoring to perhaps have some lossy behavior, in order to gain performance for point 2, as long as it conforms to the specific feature-preserving requirements from point 1."}
%\dave{I like the definition provided by Matthew above.}
To enable this capability, new algorithms such as multigrid-based hierarchical data refactoring~\cite{ainsworth2018multilevel,ainsworth2019multilevel, ainsworth2019multilevel2} have recently been developed by the applied mathematics community.
That data refactoring approach
models a dataset with a series of hierarchically organized \emph{coefficient classes}, such that an approximation of the original data with a specified fidelity can be reconstructed by using different numbers of coefficient classes.
We call the process of building coefficient classes \emph{decomposition} and the process of reconstructing data from coefficient classes \emph{recomposition}.

%{\writer{Lipeng: need a better flow here}}
\begin{figure*}[t]
    \centering
    \includegraphics[width=0.9\textwidth]{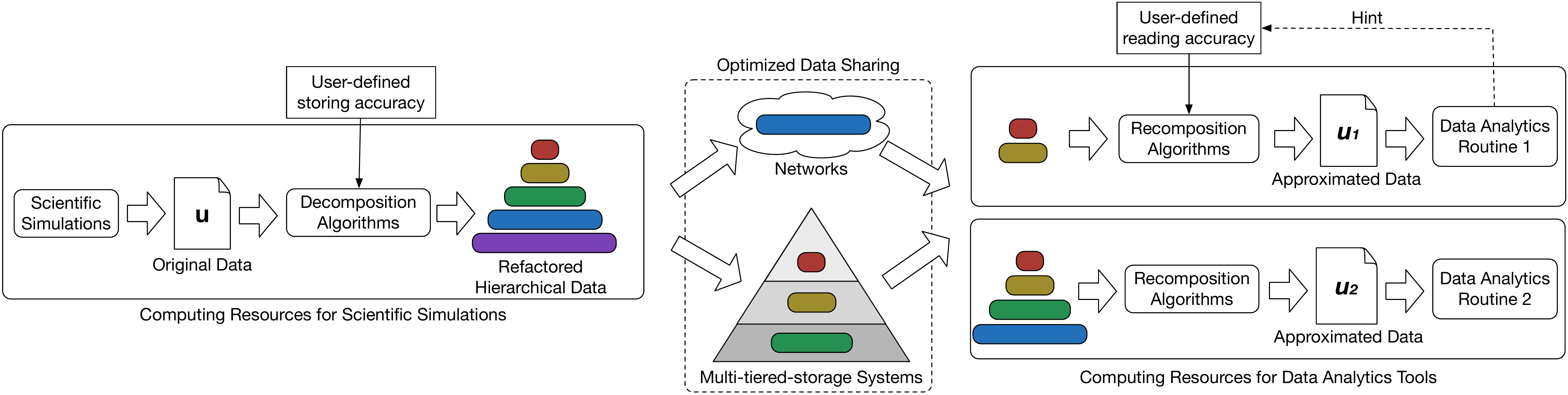}
    \caption{Example of hierarchical data refactoring helping optimize data movement in scientific workflows by intelligently moving each coefficient class across networks and/or multi-tiered-storage systems, based on available capacity and bandwidth.
    %\ian{small text is hard to read. Also, you write ``refactorizing" instead of ``refactoring' in one place.}
    }
    \label{data-refac-exp}
    \vspace{-2em}
\end{figure*}
%In this work, we explore how one can apply the capability to refactor datasets into more reasonable sizes while maintaining the fidelity of domain-driven analytics. 
%\ian{More fuzzy terms. The two terms ``more reasonable sizes'' while ``maintaining fidelity'' imply to me lossless compression, which is not what I think you are about. It would be good to define what you mean by ``more reasonable sizes": do you mean compression, or rather multigrid organization to allow rapid access to approximations? And what is the ``fidelity of domain-driven analytics"? Is the idea perhaps that some analytics do not need full resolution data?}
%\dave{Is "reasonable" a function of inputs from scientists? For example, max errors on variables or derived quantities, or expected operations on the variables,}

%where each class contains a number of coefficients, called a .
%\ian{The two terms prioritization class and coefficient class do not tell me much.}
%Approximations to the original data can be reconstructed by selecting a number of coefficient classes based on accuracy requirements.

Hierarchical data refactoring gives both data producers (e.g., scientific simulations) and consumers (e.g., data analysis routines) the flexibility to store, transport, and access data to satisfy space and/or accuracy requirements.
For example, data sharing between two coupled scientific applications~\cite{CODAR2020} can be optimized by intelligently moving coefficient classes through multi-tiered-storage systems (e.g., storage systems containing non-volatile memory, magnetic disks, and tapes) 
%\cite{Wan:2014, Wan:jpdc:2017}
and/or networks based on available capacity and bandwidth.
In Figure~\ref{data-refac-exp}, simulation data are refactored into five coefficient classes and then shared with data analysis routines via multi-tiered-storage systems and networks.
When accuracy can be estimated based on the number of selected coefficient classes, users can control the accuracy of the reconstructed data while storing and reading the data.
If user-defined accuracy requirements indicate that information encoded in the first four coefficient classes are enough for subsequent data analyses, then the fifth coefficient class can be ignored.
Then, the four coefficient classes can be intelligently shared over the storage systems and network based on their size, available bandwidth/capacities, and accuracy requirements from data analysis routines.
In the figure, Data Analysis Routine~1 needs only two coefficient classes to achieve desired accuracy, while Routine~2 needs four.
The ability to choose a reduced number of coefficient classes allows users to reduce data movement costs substantially.
%I'm not sure what the following is trying to say.
%As the data needs to be refactored or recomposed at write or read time, it is critical to ensure efficient data refactoring and recomposition processes. 

%Multigrid-based hierarchical data refactoring~\cite{ainsworth2018multilevel,ainsworth2019multilevel, ainsworth2019multilevel2,ainsworth2020multilevel} represents a promising class of data refactoring approach for handling scientific data.
%Comparing with wavelet-based data refactoring, multigrid-based approach gives users the flexibility to control either the accuracy or the storage space requirements of the reconstructed data. 

As great as the benefits of reduction in data movement and management costs may be, if the decomposition and recomposition routines are too expensive, then the total process is less useful in production.
The use of Graphics Processing Units (GPUs) has shown significant increases for scientific computations that can be adopted to the streaming execution model. These increases are due to the high parallel computational power and memory throughput in GPUs.
As the algorithms involved in multigrid-based hierarchical data refactoring are highly parallelizable, using GPUs to accelerate its routines is attractive.
Also, we anticipate that if used with merging GPU communication technologies~\cite{li2018tartan, li2019evaluating} (e.g., NVLink, GPUDirect RDMA, etc.), GPU data refactoring would be greatly beneficial for speeding up data sharing for both CPU- and GPU-based scientific applications.

We focus here on accelerating the two major routines, decomposition and recomposition, in multigrid-based data refactoring on GPUs and evaluating the benefit for producer and consumer applications.
%Specifically, we aim to accelerate both the refactoring and recomposition processes on GPUs so that they can benefit both data producer and consumer applications.
Although the multigrid-based algorithms are naturally parallelizable, achieving good performance require carefully designed parallel algorithms together with deep optimizations for GPU architectures.
Our specific contributions, and the sections in which they are described, are as follows.

In \S\ref{sec:design}, 
we describe the first multigrid-based data refactoring routines for modern GPU architectures.
Specifically, we present three optimized kernels for data refactoring on GPU. These optimizations can balance both minimizing memory footprint and improving memory access efficiency. 
Based on the optimized kernels we introduce the designs of GPU data refactoring that is optimized for spatiotemporal or high dimensional scientific data.
We also provide optimizations for consumer-class GPUs commonly used in edge systems and dense multi-GPU architectures commonly used in HPC systems.
    
In \S\ref{sec:eval}, 
we demonstrate our design by implementing the state-of-the-art non-uniform multi-dimensional multigrid-based data refactoring algorithms of Ainsworth et al.~\cite{ainsworth2018multilevel,ainsworth2019multilevel, ainsworth2019multilevel2},
and show that our methods perform well on both a consumer-class desktop and the Summit supercomputer, achieving 160$\times$ and 15$\times$ speedups compared with state-of-the-art CPUs and GPUs, and 264 TB/s throughput on 1024 Summit nodes.

In \S\ref{sec:showcase},
we use two common scenarios in scientific computing to showcase our work: 1) reducing data movement costs between simulations and in situ visualization applications; and 2) speeding up lossy compression for scientific data. % (\S\ref{sec:showcase}).

%% file: tex/backgrounds.tex
\section{Background}\label{sec:background}
\subsection{Theory of multigrid-based hierarchical data refactoring}
The multigrid-based hierarchical data refactoring developed by Ainsworth et al. support nonuniformly-spaced structured multidimensional data, commonly found in scientific computations, by using
%For the simplest case, when refactoring 3D data with dimension of $(2^L+1)^3$, the number of nodes at grid level $l$ equals to $(2^l+1)^3$ (with $L$ being the finest grid and 0 being the coarsest grid).
hierarchical representations to approximate data. Specifically, they decompose data from fine grid representation to coarse grid representation in an iterative fashion, with a global correction to account for the impact of missing grid nodes in each iteration.
% , which is used to account for the impact of the absence for the missing fine grid nodes.  
% At decomposition level $l$, we approximate the values on the nodes in $N_l \setminus N_{l-1}$ with piecewise linear interpolation and modify the values of coarse grid nodes with the global correction.  We illustrate one step of the multigrid decomposition algorithm in Figure~\ref{example-1d}, with the notation shown in Table~\ref{tab:notation}.
% starts with the finest grid resolution $(2^L + 1)\times(2^L + 1)\times(2^L + 1)$, and works toward the coarsest grid resolution (i.e., $2\times2\times2$) in an iterative fashion with grid resolution reduced to half for each iteration.
% Denote $u = \sum\sum\sum u_{ijk}\phi^L_{ijk}$ as the original function which takes value $u(i, j, k) = u_{ijk}$ for each node, where $\{\phi^L_{ijk}\}$ are nodal basis functions\footnote{def} in $V_L$. 
If we use functions to represent the discrete values continuously, the decomposition from fine grid level $l$ to coarser grid level $l-1$ can be formulated with the notation in Table~\ref{tab:notation} as follows,

%\resizebox{.45 \textwidth}{!} 
%{
\vspace{-1em}
\begin{equation}
\label{e1}
\underbrace{Q_{l-1}u}_{\substack{\text{Projection} \\ \text{onto} V_{l-1}}} = \underbrace{Q_{l}u}_{\substack{\text{Projection} \\ \text{onto} V_{l}} }- \underbrace{(I-\Pi_{l-1})Q_{l}u}_{\text{Coefficients}} +  \underbrace{(Q_{l-1}u - \Pi_{l-1}Q_{l}u)}_{\text{Corrections}}
\end{equation}
%}

\noindent
where the piecewise linear function $u$ takes the same values as the original data for each node; $Q_{l-1}u$ and $Q_lu$ are the function approximations of $u$ at levels $l-1$ and $l$, respectively; $(I-\Pi_{l-1})Q_{l}u$ is the difference between the values of the fine grid nodes at level $l$ and their corresponding piecewise linear approximations; and $(Q_{l-1}u - \Pi_{l-1}Q_lu)$ is the global correction.
% Two decomposition steps of a 1D example is illustrated in  Figure \ref{}
According to Eq.~\eqref{e1}, two major steps are involved at each level of the multigrid decomposition: 1) compute coefficients for the current multigrid level $l$; and 2) compute the global correction and add it to the nodes in the next coarse grid (level $l-1$). In what follows, we introduce how to compute coefficients and corrections.
% to match the $L^2$ projection of the original data onto the space $V_{l-1}$, the function space with respect to nodes at grid level $l-1$.  
% The next iteration starts on the coarser grid (level $l-1$) and decomposes the next coarser grid (level $l-2$).
% At each level, data that are present in the next coarser grid level, but absent in current grid level, are approximated via multilinear interpolation. 
% For example, bilinear and trilinear interpolation are used for 2D and 3D grids, respectively.
% Mathematically, the approximated data at grid level $l$ (i.e., $Q_lu$) can be decomposed into three components, as follows.

\begin{table}[t]
\centering
\caption{Notation used in algorithms, formulations, figures}
\label{tab:notation}
\begin{tabular}{|c|l|}
\hline
Symbol &  Description\\ \hline
$u$ &  Function represented by the original data.\\ \hline
$N_l$ &  Nodes at grid level $l$.\\ \hline
$C_l$ &  Coefficients at grid level $l$.\\ \hline
$V_l$ &  Function space with respect to $N_l$.\\ \hline
$Q_l$ &  The $L^2$ projection onto $V_l$.\\ \hline
$\Pi_l$ &  The piecewise linear interpolant in space $V_l$.\\ \hline
$a \rightarrow b$ & $b$ is calculated using $a$.\\ \hline
\end{tabular}
%\vspace{-0.5em}
\end{table}

\subsubsection{Compute coefficients}
\begin{figure}[t!]
    \centering
    \begin{subfigure}[t]{0.24\textwidth}
    \includegraphics[width=0.9\textwidth, trim=4mm 4mm 1mm 3mm, clip]{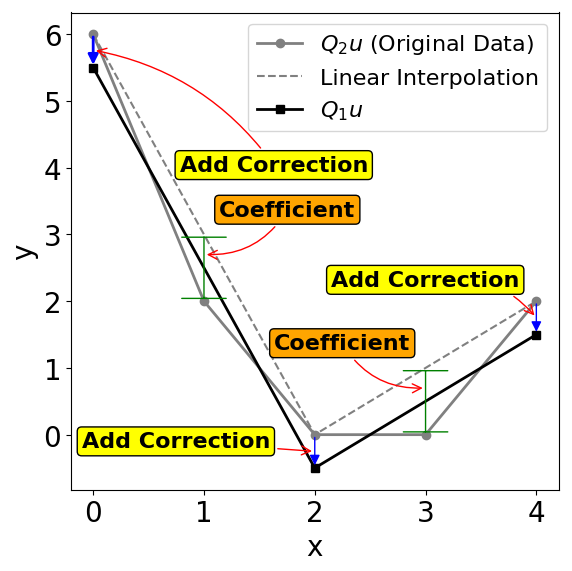}
    \vspace{-1em}
    \caption{Decomposition at $l=2$}
    \end{subfigure}
    \begin{subfigure}[t]{0.24\textwidth}
    \includegraphics[width=0.9\textwidth, trim=2mm 4mm 3mm 3mm, clip]{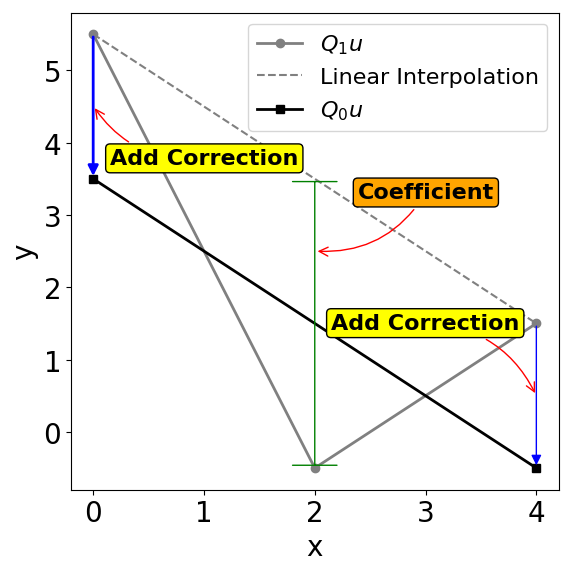}
    \vspace{-1em}
    \caption{Decomposition at $l=1$}
    \end{subfigure}
    \caption{Example of decomposing a 1D dataset produced from discretizing a quadratic function: $y = x^2-5x+6$}
    \label{example-1d}
\end{figure}

%Mathematically, 
The coefficients store the difference between the data approximated by nodes at levels $l$ (i.e., $N_l$) and $l-1$ (i.e., $N_{l-1}$) before corrections are added.
Since $N_{l-1}$ is contained in $N_l$, its nodes have the same values in both levels; thus the nonzero differences only occur on nodes in $N_l \setminus N_{l-1}$. 
Figure~\ref{example-1d} shows how coefficients are calculated along one dimension through linear interpolation.
% To extend the coefficient calculation to three or more dimensions, a combination of linear, bilinear, and trilinear interpolation is used. 
It can be generalized to multi-dimensional cases easily by using multi-linear interpolations for approximation.
%The algorithm used to calculate coefficients for 3D data is shown in Algorithm \ref{alg-coeff}.

% \begin{figure}[h]
%     \centering
%     \includegraphics[width=0.45\textwidth]{figures/coeff-exp.pdf}
%     \caption{Calculating coefficient along one dimension}
%     \label{coeff-exp}
% \end{figure}

\begin{figure*}[ht]
    \centering
    \includegraphics[width=0.8\textwidth]{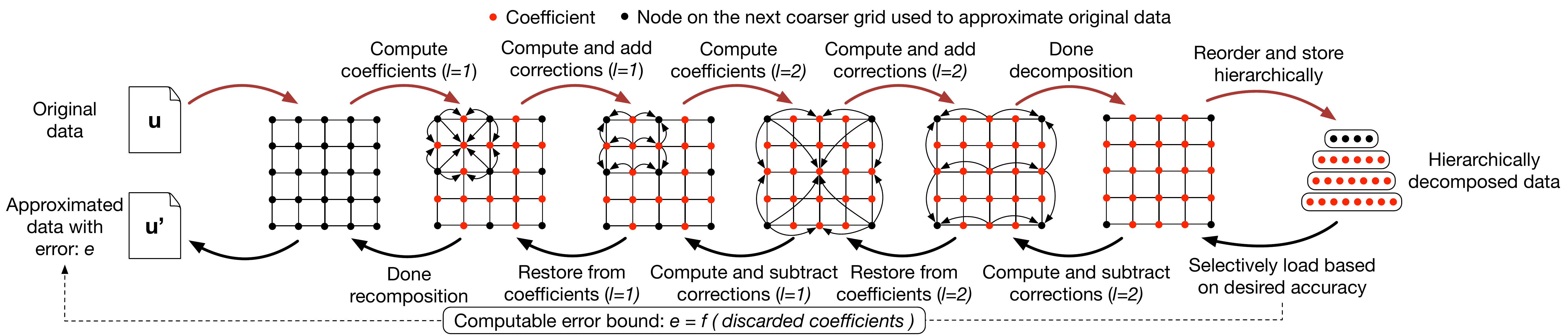}
    \caption{Multigrid-based data refactoring: Decomposition (left to right); recomposition (right to left).}
    \label{refac-recomp-example}
    \vspace{-2em}
\end{figure*}

\subsubsection{Compute correction}
%\todd{I do not understand the next two sentences.}
% To make values at the next coarser grid $l-1$ represent the $L^2$ projection onto the function space $V_{l-1}$, adjustments to each nodal value need to be made. 
% This involves calculating the correction.
Ainsworth et al.\ prove that the correction is the orthogonal projection of the calculated coefficients at grid level $l$ onto $V_{l-1}$~\cite{ainsworth2018multilevel}; thus, adding the correction to the next coarse grid better approximates data in the current grid. 
To explain, we first define $z_{l-1}$ as the correction for grid at level $l-1$. From Eq.~\eqref{e1}, we have that:

\vspace{-1em}

\begin{equation}
\label{e2}
z_{l-1} - \underbrace{(I-\Pi_{l-1})Q_{l}u}_{\text{Coefficients}} = -(Q_{l} - Q_{l-1})u \in V_{l-1}^{\bot}
\end{equation}

\vspace{-0.5em}

If we apply $L^2$ projection at grid level $l-1$ (i.e., $Q_{l-1}$) to both sides of Eq.~\ref{e2}, it leads to a zero function since it belongs to $V_{l-1}^{\bot}$. 
Also, since $z_{l-1}$ is in $V_{l-1}$, $Q_{l-1}z_{l-1} = z_{l-1}$.
So, we can see that $z_{l-1}$ is the orthogonal projection of the coefficients onto $V_{l-1}$. Namely, $Q_{l-1}(I-\Pi_{l-1})Q_{l}u = z_{l-1}$.

The correction can thus be computed by solving a variational problem: find $z_{l-1} \in V_{l-1}$ such that $(z_{l-1}, v_{l-1}) = ((I-\Pi_{l-1})Q_{l}u, v_{l-1})$ for all $z_{l-1} \in V_{l-1}$.
Then, $z_{l-1}$ can be found by solving linear systems
$M_{l-1}z_{l-1} = f_{l-1}$
where $M_{l-1}$ is a tensor product of the mass matrices~\cite{mass} of each dimension, i.e., $ M_{l-1} = M^1_{l-1} \otimes M^2_{l-1} \cdots \otimes M^d_{l-1}$, where $d$ is the number of dimensions and $f_{l-1}$ is the load vector, which can be calculated using:
$f_{l-1} =  R_lM_l \text{vec}(C_l)$,
where $R_l$ is a transfer matrix that coverts basis functions from $V_l$ to $V_{l-1}$ and $C_l$ is the coefficient matrix at level $l$, which consists of computed coefficients at $N_l \setminus N_{l-1}$ and zeros at $N_{l-1}$.
% In terms of dimensions, we take a 2D dataset for an example. 
% Assume the grid at level $l$ has $5 \times 5$ nodes and grid at level $l-1$ has $3 \times 3$ nodes. 
% Then, $C_l$ has dimensions of $5\times 5$. So, vectorized $C_l$ has dimensions of $5^2 \times 1$. $M_l$ has dimensions of $5^2 \times 5^2$. $R_l$ has dimensions of $3^2 \times 5^2$. $f_{l-1}$ has dimensions of $3^2 \times 1$. When solving $M_{l-1}z_{l-1} = f_{l-1}$, $M_{l-1}$ has dimensions of $3^2 \times 3^2$ and $z_{l-1}$ has dimensions of $3^2 \times 1$. After de-vectorizing $z_{l-1}$, the correction has dimensions of $3 \times 3$, which is added to the nodes at level $l-1$.

\subsubsection*{Overall decomposition/recomposition process}
Figure~\ref{refac-recomp-example} illustrates this process 
on a 5$\times$5 2D dataset. 
The original data is on the left, and the refactored representation is on the right.
The decomposition process moves from left to right (i.e., from finest to coarsest grid) and involves four steps: computing coefficient and computing correction (II.A.1 and II.A.2) for each of the two levels.
For multi-dimensional data, the computation of correction is done by working on each dimension in a prescribed order~\cite{liang2020optimizing}; in this 2D example, it proceeds first along the rows and then along the columns.
Recomposition moves from right to the left: i.e., from coarsest to finest grid.
There are again four total stages, but these occur in the reverse order.
The approximation of the original data is produced after recomposition.
Based on how coefficients are omitted in recomposition, an error bound on data approximation can be computed~\cite{ainsworth2019multilevel}.
% First, the corrections are calculated using coefficients and used to undo the corrections (i.e., subtract values) on the next coarser grid to restore their nodal values. 
% Then, the coefficients are restored to their original values by adding the interpolated values from the next coarser grid.

\subsection{Existing GPU-based data refactoring} 
The state-of-the-art MGARD~\cite{mgard} GPU-based data refactoring system
redesigns original serial algorithms to expose high parallelism to suit the many-core architecture of modern GPUs.
It achieves $O(n^3)$ thread concurrency for computing coefficients and $O(n^2)$ thread concurrency for computing corrections, and applies node reordering such that each kernel can take advantage of coalesced memory accesses. 
Theoretically, with large inputs, these levels of thread concurrency are more than enough to fully occupy GPU cores that can help achieve high data refactoring throughput. 
However, performance evaluation shows that it still suffers from underutilized memory throughput, achieving less than 10\% of theoretical peak.

%% file: tex/design.tex
\section{Designing GPU-accelerated data refactoring}
\label{sec:design}
We next discuss the design of our GPU-accelerated multigrid-based hierarchical data refactoring method. 
We first focus on the optimizations for each computing kernel involved in data refactoring.
Then, we discuss how to use heuristic auto tuning to maximize the refactoring throughput.
Finally, we provide overall designs of spatiotemporal refactoring algorithms with optimizations.
% We classify the computing patterns into three categories and propose three general kernel designs for GPUs.
% Following the efficient kernel designs, we discuss optimizations to help each of the kernels efficiently work together so that their performance can be maximized.
% Finally, we discuss design details about how to use heuristic auto tuning to maximize the refactoring throughput.

\subsection{Designing optimized GPU multigrid kernels}
Decomposition and recomposition each involve three major 
%computational 
steps: 1) computing coefficients; 2) mass-transfer matrix multiplication; and 3) correction solver.
Based on their computation pattern, we can classify them into three categories: \textit{grid processing style}; \textit{linear processing style};  and \textit{iterative processing style}. 
We design kernels dedicated for each processing style. 

\subsubsection{Grid processing kernel (\gpk)}
Grid processing style has the characteristic of processing data grid-wise.
Namely, it processes nodes within the domain of a grid in a certain resolution level (e.g., $N_l$) or between neighboring levels (e.g., $N_l$ and $N_{l-1}$).
In the multigrid-based data refactoring, the calculation of coefficients follows the grid processing style.
The major calculation is to compute the interpolation at nodes in $N_l \setminus N_{l-1}$ using nodal values in $N_l$. 
Parallelization can favor either interpolation operations (i.e., parallelism $\propto O(N_l \setminus N_{l-1})$) or accessing nodal values (i.e., parallelism $\propto O(N_{l})$).  The former can lead to a less thread divergence, while the latter can achieve a higher memory access efficiency.
The computation of coefficients is a memory bound operation, as its time complexity is $O(n)$.
Therefore, it is essential to optimize in favor of memory access efficiency instead of computation.
This is also chosen in the state-of-the-art GPU data refactoring~\cite{mgard}.

\begin{figure}[ht]
    \vspace{-0em}
    \centering
    \includegraphics[width=0.7\columnwidth]{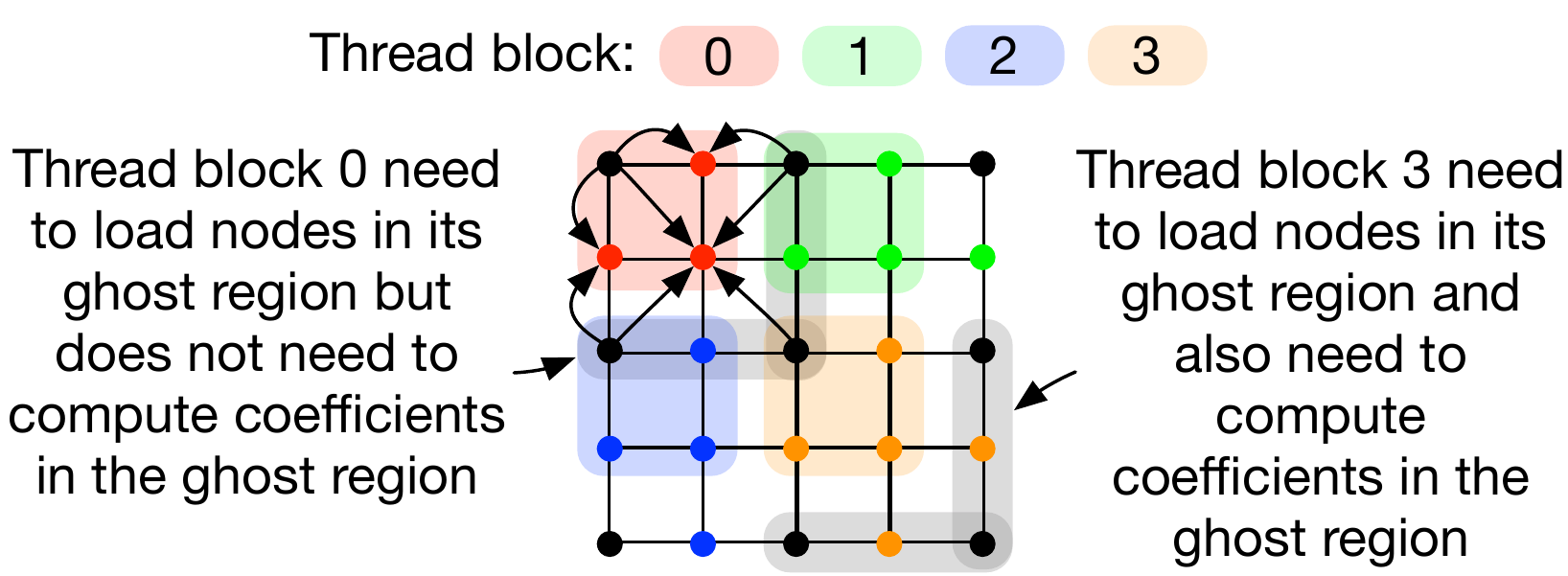}
    \caption{Workload of computing coefficient is distributed among 4 thread blocks. Calculating coefficients in corresponding ghost regions is needed for some thread blocks (e.g., thread block 1, 2, and 3). }
    \label{gpk-ghost}
\end{figure}

The key strategy they used to optimize for memory access is to use shared memory to cache a block of data for process, of which the nodes values are loaded/stored in a coalesced-friendly fashion.
However, we identify that keeping efficient data movement on memory bound computation is not enough to achieve good performance. 
The level of thread divergence in computation can still have a great impact on the overall performance and sometimes it can wrongly convert computation from memory bound to compute bound.
The reason is threefold: 1) high degrees of thread divergence can great increase the total cycles cost in computation; 2) variable floating point operation counts caused by different interpolation types further brings workload imbalance which leads to longer idling cycles; 3) as shown in Figure~\ref{gpk-ghost}, some thread blocks also need to calculate coefficients in the ghost region, which exacerbate the effect of thread divergence.

However, we found that keeping efficient memory access patterns is not exclusive with having low thread divergence.
In designing our \gpk, we propose to decouple memory access and computation on nodal values in terms of thread-node assignment through a thread reassignment strategy.
Specifically, we use two different thread-node assignment for loading/storing nodal values and computing interpolations such that we keep maintaining efficient coalesced memory access pattern while having one warp process the same type of interpolations along the same dimension.

\begin{figure}[ht]
    \centering
    \vspace{-1em}
    \includegraphics[width=0.8\columnwidth,trim=2.5mm 2mm 0mm 1mm,clip]{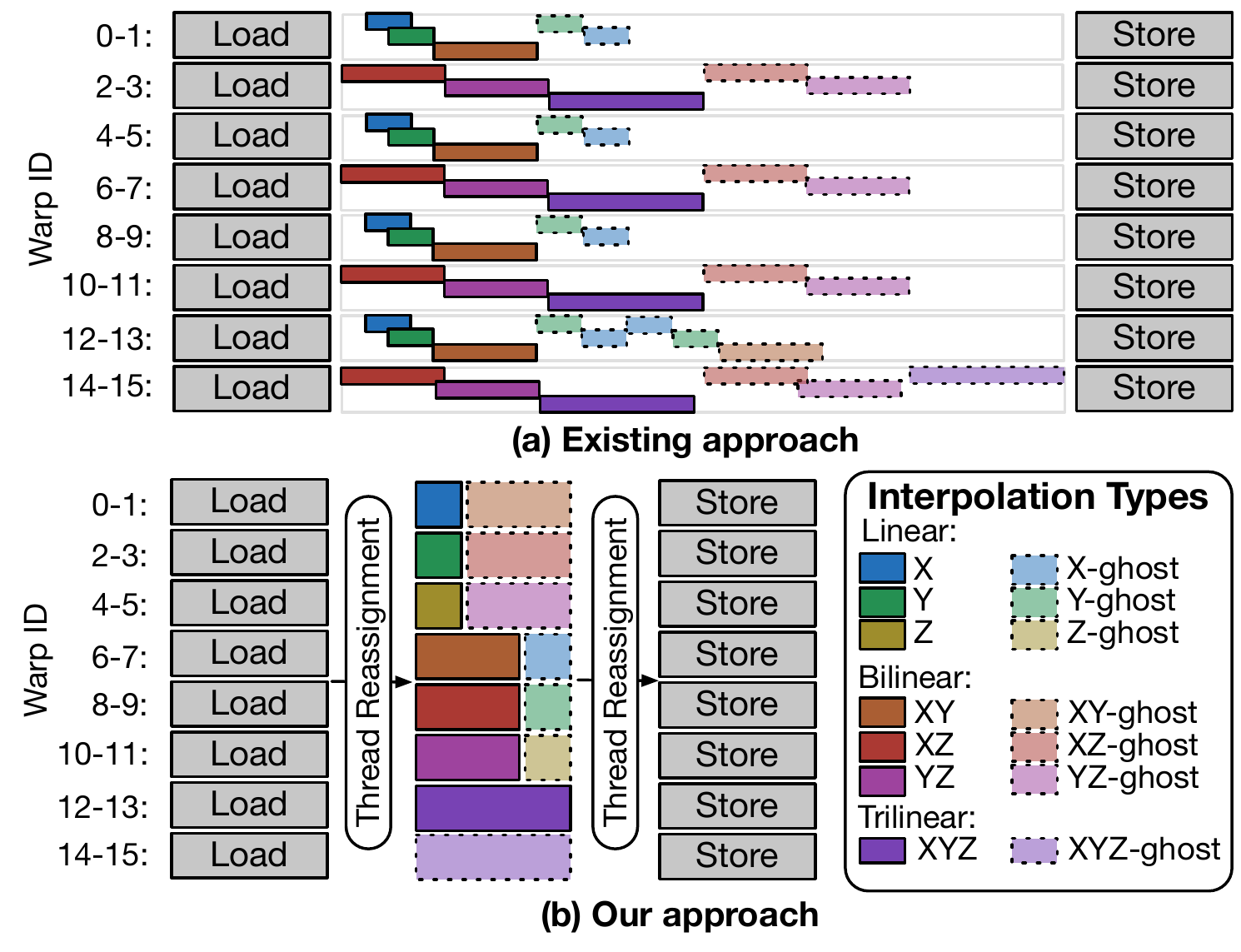}
    \caption{Conceptual flow of a thread block with $8\times8\times8$ threads (16 warps) calculating coefficients using existing and our grid processing kernel \gpk. The thread reassignment strategy allows \gpk~to greatly reduce thread divergence. }
    \label{gpk}
    \vspace{-0em}
\end{figure}

Figure~\ref{gpk} shows the conceptual execution flow of one thread block using existing approach and our proposed \gpk~when computing coefficients.
As nodes in $N_{l}$ need to be shared with neighbors during interpolation operations,  we let each thread block coordinate work on a block of data and use shared memory as a scratch space. 
We organize threads such that threads in the same warp load values that are consecutive in memory to achieve efficient coalesced memory access patterns.
For computing, we apply a thread re-assignment strategy to achieve divergence-free execution.
% Algorithm \ref{alg-coeff} shows how we calculate the thread-interpolation operation assignments that minimizes thread divergence.
% It is easy to see that the reassignment processing brings negligible computing overhead. 

\subsubsection{Linear processing kernel (\lpk)} 
\begin{figure}[ht]
    \vspace{-1em}
    \centering
    \includegraphics[width=0.5\textwidth]{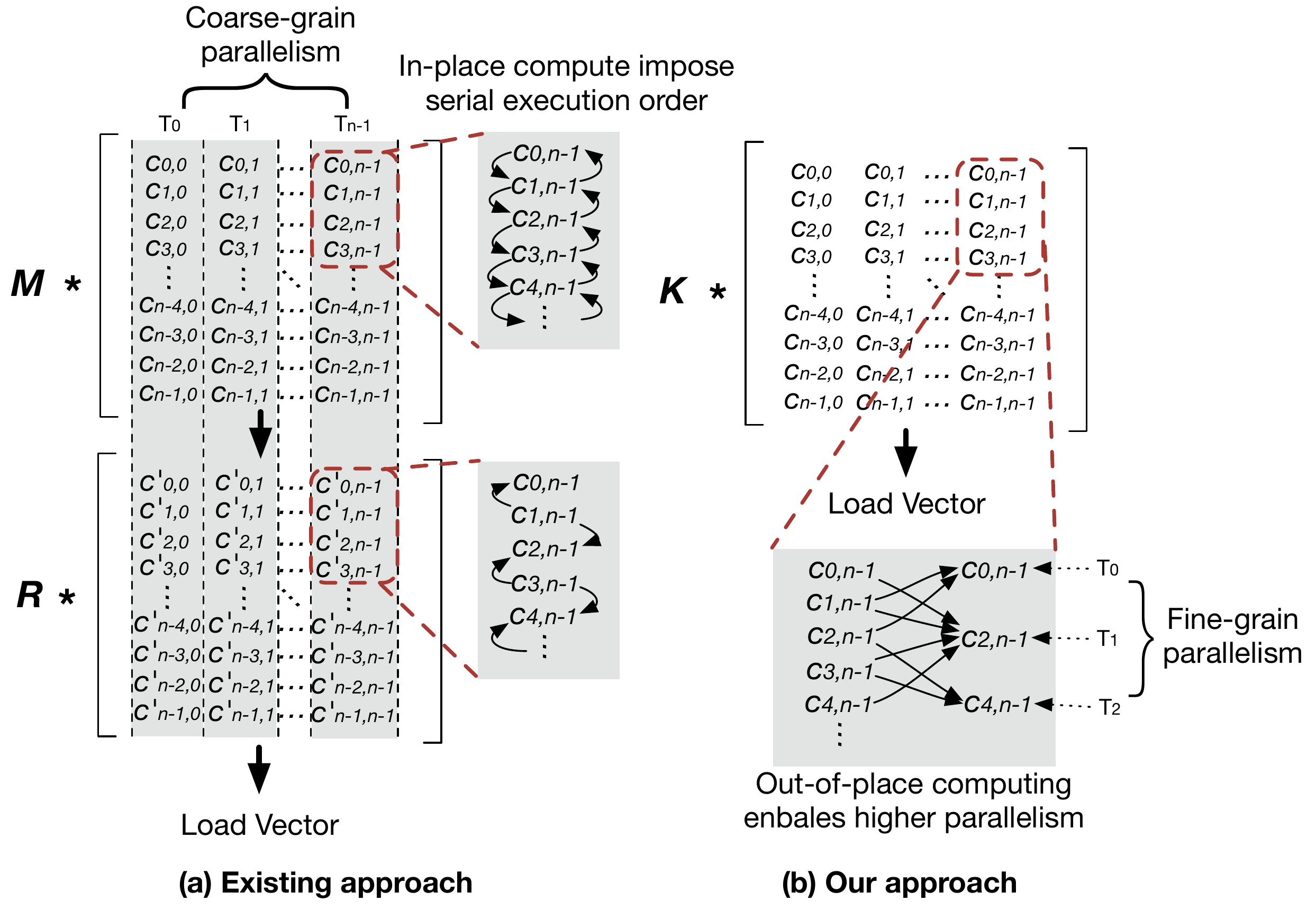}
    \caption{The conceptual workflow of mass and transfer multiplication using existing approach and proposed approach. Through optimizations, our approach achieves finer grain parallelism.}
    \label{lpk}
    \vspace{-1em}
\end{figure}

The linear processing style computes stencil operations on elements in vectors along one dimension in a grid.
In multigrid-based data refactoring, when multiplying the mass and transfer matrices with computed coefficients, the computations become stencil operations, as the matrices are defined as:
\[
M_{ij}=\left\{\begin{matrix}
2(h_{i}+h_{i+1}) & \text{if~} i=j\\ 
h_{i} & \text{if~} |i-j|=1\\ 
0 & \text{else} 
\end{matrix}\right.
\]
\vspace{-0.5em}
\[
R_{ij}=\left\{\begin{matrix}
1 & \text{if~} i=j/2\\ 
r_{j-1} & \text{if~} i=(j-1)/2\\ 
1-r_{j} & \text{if~} i=(j+1)/2 \\
0 & \text{else} 
\end{matrix}\right.
\]
where $h_{i}$ is the spacing between the $i^{th}$ node, and the ${i+1}^{th}$ node and $r_i = h_{i}/(h_{i}+h_{i+1})$.
As shown in Figure~\ref{lpk}(a), each value of each node needs to be computed using the original values of its neighbors, which means it cannot update its stored value unless all neighbors have finishing using its original value for computation.
Such data dependencies present a dilemma for kernel design: common out-of-place designs (i.e., element-wise parallelism) bring high parallelism but also high memory footprint; on the other hand, in-place design (i.e., vector-wise parallelism), used in \cite{mgard}, sacrifices the opportunity to exploit intrinsic parallelism.

To eliminate this dilemma, we design a novel linear processing kernel (\lpk) with four optimizations.
First, we change the original computation from in-place to out-of-place to achieve finer-grain parallelism.
Second, we merge the mass and transfer matrices to reduce computational costs.
We call the new matrix \texttt{mass-trans}, which is defined as:
$$K_{ij}=\left\{\begin{matrix}
(2+r_{j-2})h_{j-1}+(1+r_{j}) & \text{if~} i=j/2\\ 
(2r_{j-2}+1)h_{j-1}+2r_{j-2}h_{j-2} & \text{if~} i=(j-1)/2\\ 
(3-2r_{j})h_{j+1}+2r_{j+1}h_{j+1} & \text{if~} i=(j+1)/2 \\
r_{j-2}h_{j-2} & \text{if~} i=(j-2)/2\\ 
(1-r_{j})h_{j+1} & \text{if~} i=(j+2)/2\\ 
0 & \text{else} 
\end{matrix}\right.$$
Third, we use shared memory to cache a tile of nodes to allow sharing of coefficients (input) between different threads, so as to reduce total accesses to global memory.
Finally, to reduce extra memory footprint we use kernel fusion technique to fuse the operation of copying coefficients with the multiplication of the mass-trans matrix with coefficients along the first dimension.
By eliminating the need of storing a copy of the computed coefficient in the workspace, we avoid large increase in the overall memory footprint.

\subsubsection{Iterative processing kernel (\ipk)}
The iterative processing style has the characteristic of processing nodes in a grid that contains strong data dependencies such that nodes have to be processed iteratively in a certain order. 
In multigrid-based data refactoring, the correction solver needs to solve for the corrections.
We use the Thomas algorithm~\cite{atkinson1985elementary}, which needs a forward and a backward pass on the load vector. 
Since the load vectors along one dimension can be solved independently, they can be solved in parallel.
This level of parallelization is well exploited in \cite{mgard}.
Specifically, they assign each thread to handle the solving process of one load vector independently.
Although this brings high thread concurrency with divergence free execution, it actually suffers from inefficient memory accesses for two reasons: first when solving vectors on leading dimension full coalesced memory access cannot be achieved (actual achieve efficiencies are about only 12\% and 25\% for single and double precision data); second, compared with \gpk~and \lpk, \ipk~only has $O(n^2)$ degrees of thread concurrency, which may brings less on-the-fly memory accesses to fully utilizes the memory bandwidth. 

\begin{figure}[ht]
    \vspace{-1em}
    \centering
    \includegraphics[width=0.8\columnwidth,trim=2mm 0mm 2mm 2mm,clip]{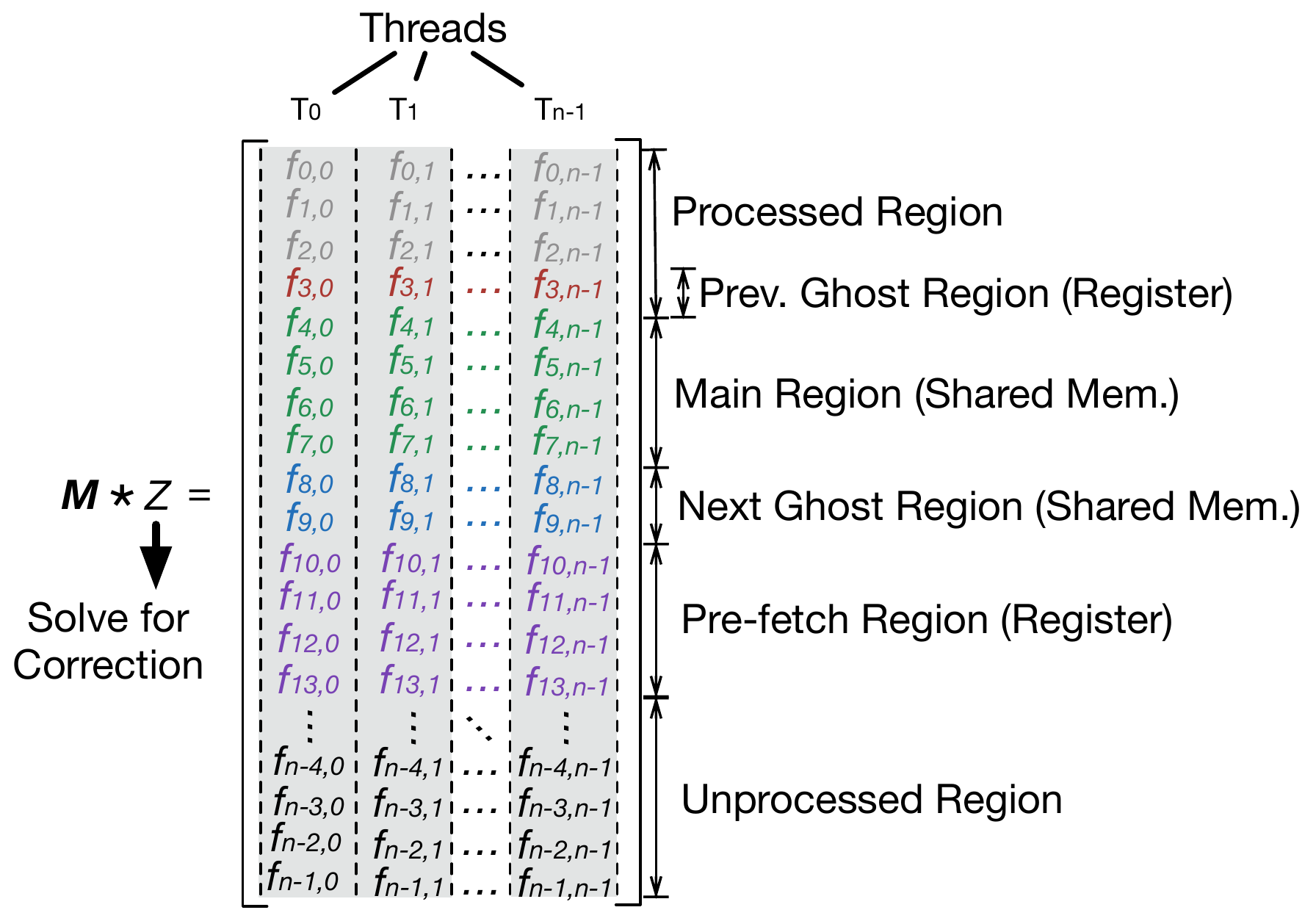}
    \caption{Correction solver designed following iterative processing kernel (\ipk). The node vectors are partitioned into six regions during processing. The use of shared memory ensure efficient  coalesced memory accesses regardless of which dimension it is processing. Data prefetching further increases concurrency on memory accesses. }
    \label{mass-exp}
    \vspace{-2.5em}
\end{figure}

To address this issue, we proposed a novel processing kernel, \ipk, that can guarantee efficient coalesced memory access patterns with high concurrent memory accesses.
We first parallelize the vectors by assigning a batch to a thread block.
Since the update of each node depends on its neighboring elements, we use shared memory as scratch space to avoid polluting the un-processed nodes.
Specifically, we let each thread block iteratively work on a segment of load vectors at a time until the whole vector is updated.
Thus, as shown in Figure~\ref{mass-exp}, during the computation we divide the elements in the vectors into six regions: 1) the processed region stores updated elements (gray); 2) the main region consists of elements that the current iteration is working on (green); 3) and 4) due to dependence on the neighboring elements, the original value of elements in the two ghost regions (red and cyan) are needed to update the elements in the main region; 5) for better streaming processor utilization, we pre-fetch data needed for the next iteration (purple); and 6) we mark the unprocessed region as in block.
The regions move forward as the computation proceeds.
One challenge to design the algorithm is to simultaneously consider maximizing coalesced global memory access patterns, minimize bank conflict in accessing shared memory, and minimize thread divergence.
We use a dynamic data-thread assignment strategy~\cite{chen2019tsm2,rivera2021tsm2x, chen2016online, chen2018fault, chen2016gpu} to optimize both the accessing and computation of coefficients.

\subsection{Heuristic Performance Auto Tuning}
When launching each proposed kernel, choosing the execution parameters is important for its achieving good performance, since even with optimized design they can still greatly impact the efficiency of memory accesses, warp divergence, context switch overhead, etc.
Auto tuning technique is an effective approach for searching the optimum configurations.
However, brutal force search can be expensive and thus impractical.
Thus, we propose to use a heuristic auto tuning approach guided by theoretical performance models for our GPU data refactoring.
We first build performance models for the three kernels we proposed.
Among all tunable execution parameters, we find that the size of the thread block ($B_x, B_y, B_z$) plays an important role in determine each kernel's performance.
Since we eliminate majority of the thread divergence and inefficient computations, we assume the memory load/store take the majority time, so we only consider total amount of memory transactions with their efficiency.
The estimated execution time of each kernel can be modeled as:
\begin{align*}
T_{\textrm{GPK}} = & \ceil{B_x+1/(S/L)} \cdot (S/L) \cdot (B_y+1) \cdot (B_z+1)\ \cdot \\
&\floor{N/B_x} \cdot \floor{N/B_y} \cdot \floor{N/B_z} \cdot 2 L\ \cdot \\
& (1/\text{Peak Mem. Band.})
\end{align*}
\begin{align*}
T_{\textrm{LPK}} = & \left(\ceil{B_x/(S/L)} \cdot S/L + 2 S/L \right) \cdot B_y \cdot B_z\ \cdot \\
& \floor{N/B_x} \cdot \floor{N/B_y} \cdot \floor{N/B_z} \cdot 2 L\ \cdot \\
& (1/\text{Peak Mem. Band.})
\end{align*}
\begin{align*}
T_{\textrm{IPK}} = & \left(\ceil{G/(S/L)} \cdot S/L + \ceil{B_x/(S/L)} \cdot S/L \cdot \ceil{N/B_x}\right)\  \cdot\\
& B_y \cdot B_z \cdot \floor{N/B_y}\ \cdot \floor{N/B_z} \cdot 2 L\  \cdot \\
&  (1/\text{Peak Mem. Band.})
\end{align*}
where $S$ is the number of bytes per memory transaction,
(32 in our test GPU); %to set it to 32 bytes.
$L$ is bytes per float (4 for single, 8 for double); and
$G$ is the dimension of the next ghost region, set to $S/L$ so that ghost data can fit into exactly one memory transaction and do not consume too much shared memory. 
Table~\ref{auto-tuning} shows the ranking of estimated performance using seven typical thread block size configurations. 
Numbers in red represent the actual best configuration as determined by profiling.
We can see our performance model can help up predict relationship between different configuration in terms of performance with relative high accuracy.
It helps us narrow down the searching space for auto tuning.
For instance, in our following evaluation, we only let the auto tuning search and pick among the estimated top three configurations to save time.

\begin{table}[]
\centering
\caption{Ranking of estimated performance of seven typical thread block size configurations; actual best in red. }
\label{auto-tuning}
\begin{tabular}{|c|c|c|c|c|c|}
\hline
$B_z$ & $B_y$ & $B_x$   & GPK & LPK & IPK \\ \hline
2 & 2 & 2   & 7   & 7   & 7   \\ \hline
4 & 4 & 4   & 6   & 6   & 1   \\ \hline
4 & 4 & 8   & 4   & 5   & 2   \\ \hline
4 & 4 & 16  & \textcolor{red}{\textbf{2}}   & 4   & \textcolor{red}{\textbf{3}}   \\ \hline
4 & 4 & 32  & 1   & 3   & 4   \\ \hline
2 & 2 & 64  & 5   & 2   & 5   \\ \hline
2 & 2 & 128 & 3   & \textcolor{red}{\textbf{1}}   & 6   \\ \hline
\end{tabular}
\end{table}

\begin{figure}[h]
    \vspace{-1em}
    \centering
    \includegraphics[width=1\columnwidth]{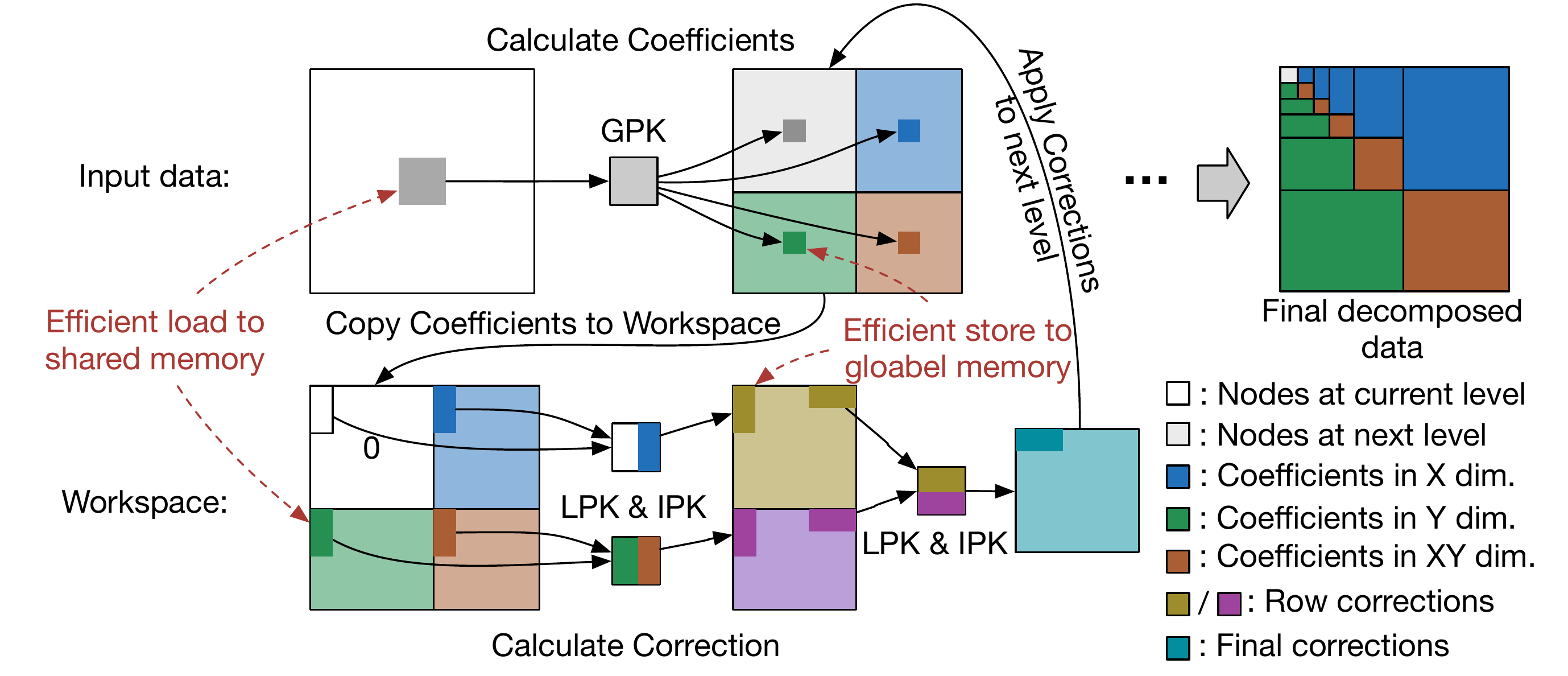}
    \caption{Decomposition with improved reordered data layout}
    \label{fig:overall}
    \vspace{-1em}
\end{figure}

\subsection{Overall refactoring process with improved ordering data layout}
Figure \ref{fig:overall} shows the design of overall GPU data refactoring routines on GPUs with improved data layout.
As computed coefficients are used for calculating corrections, which involves altering the values of coefficients.
So, to preserve the value of previously computed coefficients, the correction is computed in a workspace.
In the state-of-the-art design~\cite{mgard}, the computed coefficients are first copied to the workspace before they are used for computing corrections, which prohibits out-of-place computing unless additional memory space is used.
Our previous work leverages kernel fusion~\cite{chen2020accelerating} to merges the copy of the coefficients with the first mass-trans matrix multiplication, so that it enables us to use \lpk{} out-of-place compute without a significant increase in memory footprint.
However, one drawback with our previous design is the degraded memory access efficiency for \gpk{} since accessing nodes in coarser grids leads to larger strided memory accesses.
So, in this work, we propose to use an improved data layout to keep the stride always equal to one without bringing extra computation or memory footprint.
As shown in figure \ref{fig:overall}, by separating nodes in the next coarser grid with coefficient, we can always keep nodes compacted.
We build this reordering processing in the data store part of \gpk{}, so that it does not bring any overhead. 
% We further extend out-of-place mass-trans matrix multiplication for processing other dimensions, which improves parallelism with slight increase in memory footprint for the workspace.
% In the state-of-the-art design, the workspace is of size $m\times n \times k$ and in our design its size is $(m+1)\times (n+1) \times k$, assuming $m$, $n$, and $k$ are the three dimensions of the input data.
% The recomposition process is the opposite so we do not show its process due to page limit. 

\begin{figure}[t]
    \centering
    \includegraphics[width=0.5\textwidth, trim=0mm 0mm 0mm 0mm, clip]{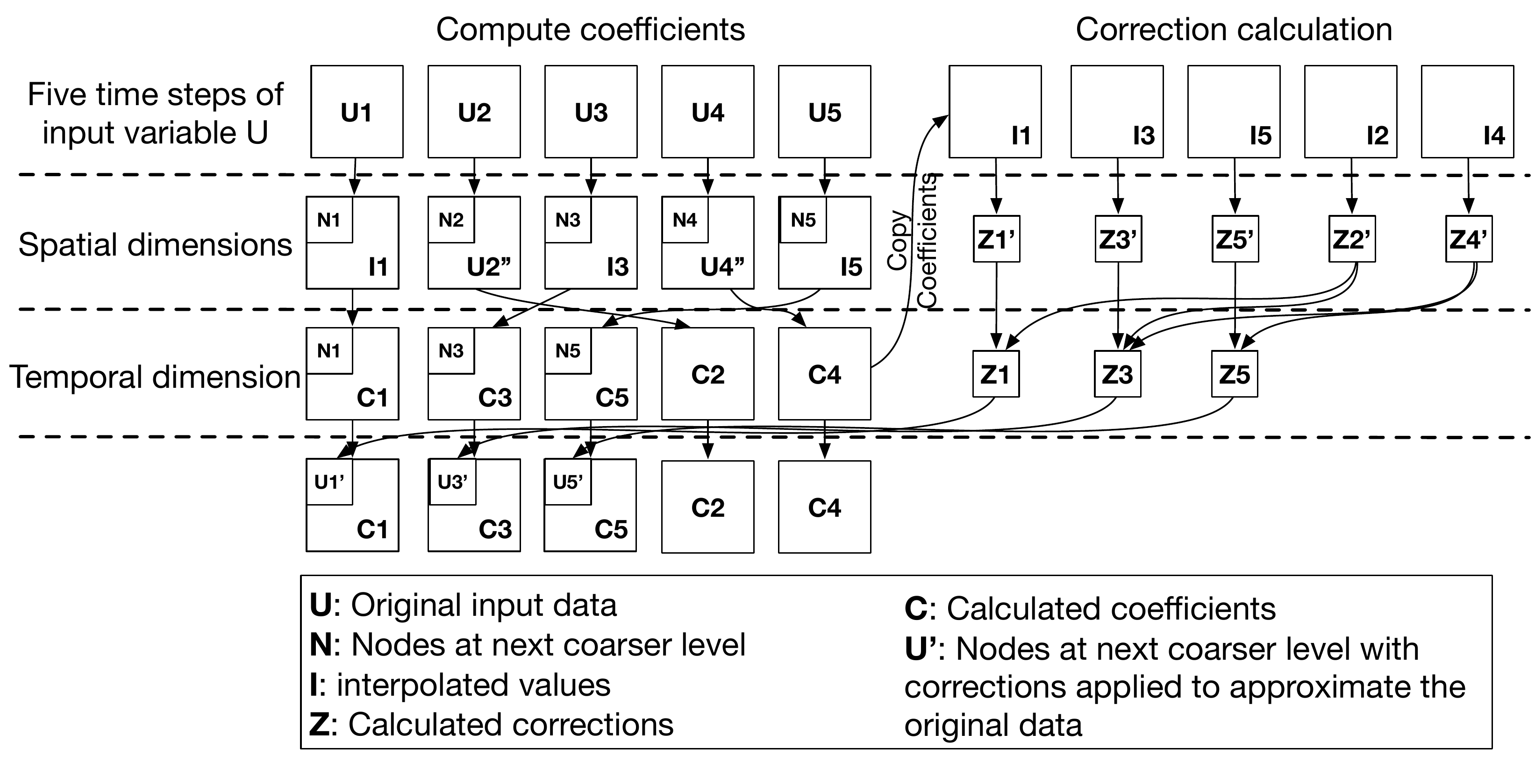}
    \caption{Illustration of one level of spatiotemporal data refactoring of a variable of five time steps}
    \label{temporal-refactoring}
\end{figure}
\subsection{Design of spatiotemporal data refactoring}
As scientific data generated from simulations or experiments commonly exhibits correlation across time, exploiting the temporal dimension in addition to the spatial dimensions can enable building a data hierarchical representation across both spatial and temporal dimensions and can potentially enable higher compression ratios when used for lossy compression.
So, it is desirable to have an optimized design for spatiotemporal data refactoring for GPUs.
Here we consider spatiotemporal data refactoring as $N+1$ dimensions of data refactoring, where $N$ is the number of spatial dimensions and $1$ is the temporal dimension.
The theory of the multigrid-based data refactoring method~\cite{ainsworth2018multilevel} can be generalized to refactor data in an arbitrary number of dimensions.
However, it is not straightforward to enable efficient GPU data refactoring on data with an arbitrary number of dimensions.

\subsubsection{Hierarchical batch optimization}
\begin{figure}[t]
    \centering
    \vspace{-1em}
    \includegraphics[width=0.4\textwidth, trim=0mm 0mm 0mm 0mm, clip]{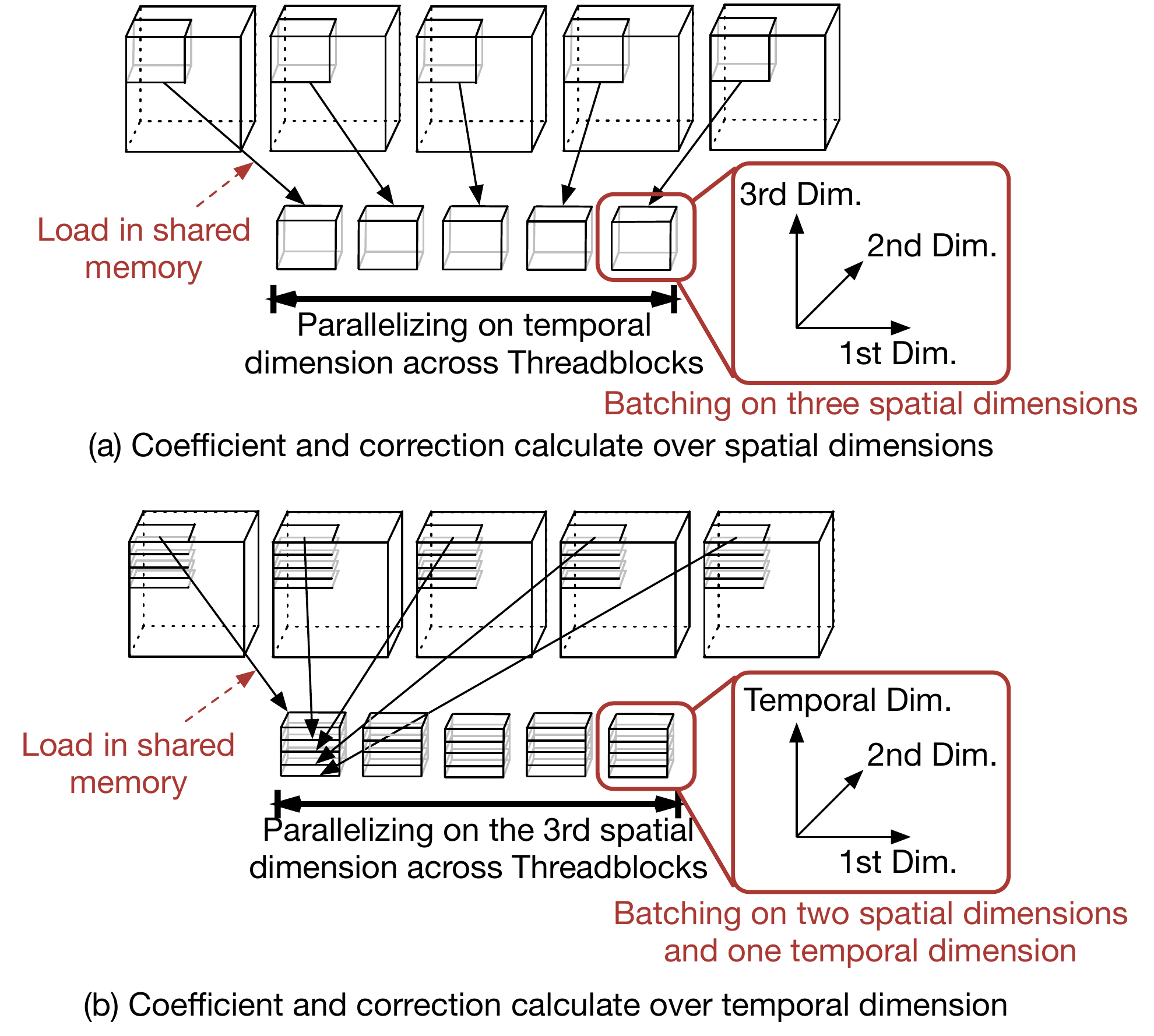}
    \caption{Hierarchical batch optimization}
    \label{batch-opt}
    %\vspace{-2em}
\end{figure}

We propose a novel design for efficient high dimensional GPU data refactoring: hierarchical batch optimization.
Namely, our hierarchical batch optimization involves two levels of batch optimization: 1) locality batch optimization; and 2) dimensional batch optimization.
Batch optimization exploiting data locality in data refactoring on GPUs was proposed in our previous work~\cite{chen2020accelerating}, which is well adopted in \gpk{}, \lpk{}, and \ipk{}.
% The main idea is to organize threads such that threads in the same warp load values that are consecutive in memory to achieve efficient coalesced memory access patterns and use shared memory as scratch space.
However, it is challenging to use locality batch optimization beyond three-dimensional data refactoring due to limited share memory space on GPUs (with batch size $b$ and total number of dimensions $d$, the required shared memory is $O(b^d)$).
Although it is possible to reduce the batch size to decrease the pressure on shared memory, that would also reduce the thread concurrency on GPU, which in turn can negatively impact performance.
So, we propose dimensional batch optimization.
Specifically, we restrict the use of shared memory to $O(b^3)$ and adaptively batch three selected dimensions.
We always keep the first spatial dimension as the selected dimension to maintain efficient memory access.
For other non-selected dimensions, we further parallel them using Threadblocks.
This is especially useful to minimize the negative impact on concurrency when the input is small in the three selected dimensions.
Figure~\ref{batch-opt} shows an example of hierarchical batch optimization for $3+1$ data.
When we process the spatial dimensions, we batch the three spatial dimensions in shared memory can parallelize the temporal dimension using multiple Threadblocks.
When we process the temporal dimension, we batch the first two spatial dimensions plus the temporal dimension in share memory and parallelize the third spatial dimension using multiple Threadblocks.
Figure~\ref{temporal-refactoring} shows our design of the overall spatiotemporal data refactoring.
We separate the processing of spatial dimensions from the temporal dimension.
Note that to improve memory access efficiency, we also extend the reordered data layout to the temporal dimension.

\subsection{Optimization for consumer-class GPUs}
Although to common sense reducing the cost of arithmetic operations has less effect for improving memory-bound computations, we find that is not always true.
This is due to the fact that not all GPUs have similar memory bandwidth and computational capacity ratios.
Consumer class GPUs are commonly used by scientists especially in the edge systems where data refactoring is particularly useful.
However, those GPUs typically offer decent memory bandwidth but limited computational power on high precision floating point data and certain operations such as division on double floating-point numbers, so memory-bound computations can be wrongly converted to compute bound including data refactoring.
We find that the fused multiply-and-add (FMA) instructions are typically efficient with high throughput even on high precision data, so we re-implement core calculation in multigrid-based data refactoring in FMA instruction as shown in Table~\ref{fma}, in which $diag(i)$ and $subdiag(i)$ are the $i^{th}$ elements of the tridiagonal mass matrix.
With this optimization, we convert more than 90\% of the floating-point arithmetic operations involved in data refactoring to use FMA.

\begin{table}[ht]
\caption{Core calculations implemented in more efficient FMA instructions}
\label{fma}
\begin{tabular}{|p{2.5cm}|p{5.5cm}|}
\hline
Operations           & Implementation in FMA instruction \\ \hline
Linear Interpolation       & result = \_\_fma($r_i$, $v_{i+1}$, \_\_fma($-r_i$, $v_i$, $v_i$))                      \\ \hline
Bilinear Interpolation.     & \makecell[l]{$t_1$ = Linear($v_{i,j}$, $v_{i,j+1}$), $r^x_{i}$)\\ 
$t_2$ = Linear($v_{i+1,j}$, $v_{i+1,j+1}$, $r^x_{i}$)\\
result = \_\_fma($t_1$, $t_2$, $r^y_{i}$)}                    \\ \hline
Trilinear Interpolation.    & \makecell[l]{$t_1$ = Bilinear($v_{i,j,k}$, $v_{i,j,k+1}$, $v_{i,j+1,k}$, \\ $v_{i,j+1,k+1}$, $r^x_{i}$, $r^y_{i}$)\\ 
$t_2$ = Bilinear($v_{i+1,j,k}$, $v_{i+1,j,k+1}$, $v_{i+1,j+1,k}$, \\ $v_{i+1,j+1,k+1}$, $r^x_{i}$, $r^y_{i}$)\\
result = \_\_fma($t_1$, $t_2$, $r^z_{i}$)}                   \\ \hline
Mass-Trans Matrix Multiplication         & \makecell[l]{$t_1$ = \_\_fma(2, \_\_fma($v_i$, $h_{i-1}$, $v_{i-2}*h_{i-2}$), \\ \_\_fma($v_{i-1}$, $h_{i-1}$, $v_{i-1}*h_{i-2}$)) \\  
$t_2$ = \_\_fma(2, \_\_fma($v_{i+1}$, $h_{i}$, $v_{i-1}*h_{i-1}$), \\ \_\_fma($v_{i}$, $h_{i}$, $v_{i}*h_{i-1}$))\\
$t_3$ = \_\_fma(2, \_\_fma($v_{i}$, $h_{i+1}$, $v_{i+2}*h_{i}$), \\ \_\_fma($v_{i+1}$, $h_{i+1}$, $v_{i+1}*h_{i}$))\\
result = \_\_fma($t_3$, $1-r_i$, \_\_fma($t_1$, $r_{i-2}$, $t_2$))}
\\ \hline
Solv. Corr. Forward. &  result = \_\_fma($v_{i-1}$, $diag(i)$, $v_{i}$)                     \\ \hline
Solv. Corr. Backward &  result = \_\_fma($-h_i$, $v_{i-1}$, $v_{i}$) * $subdiag(i)$                      \\ \hline
\end{tabular}
\end{table}

\subsection{Designing data refactoring for dense multi-GPU systems}
As dense multi-GPU systems are commonly used in modern HPC systems, we extend the GPU data refactoring~\cite{chen2020accelerating} to take advantage of multi-GPU systems.
\emph{Dense multi-GPU systems}
%~\cite{li2018tartan,li2019evaluating} 
refers to systems where multiple GPUs are equipped in each single computing node of supercomputers with fast interconnect such as NVLiks or PCIe.
There are two ways of taking advantage of multi-GPU systems: 
1) embarrassing parallel: data are partitioned refactorized independently; 2) cooperatively parallel: data are treated as a whole with multiple GPUs work on one refactoring process cooperatively.
Comparing between the two, embarrassingly parallel approach potentially have better performance as it does not bring extra communication cost. 
The cooperative parallel approach leverages data correlation information between partitions, so it potentially can have deeper hierarchical levels for refactored data and higher compression ratios when used in lossy compression.
Since the embarrassingly parallel is relatively straightforward, we mainly discuss the design of the cooperative parallel approach in this section.
We discuss the cooperative parallel design for each data refactoring kernel separately.
\begin{figure}[h!]
    \centering
    \vspace{-1em}
    \includegraphics[width=0.25\textwidth, trim=0mm 30mm 30mm 0mm, clip]{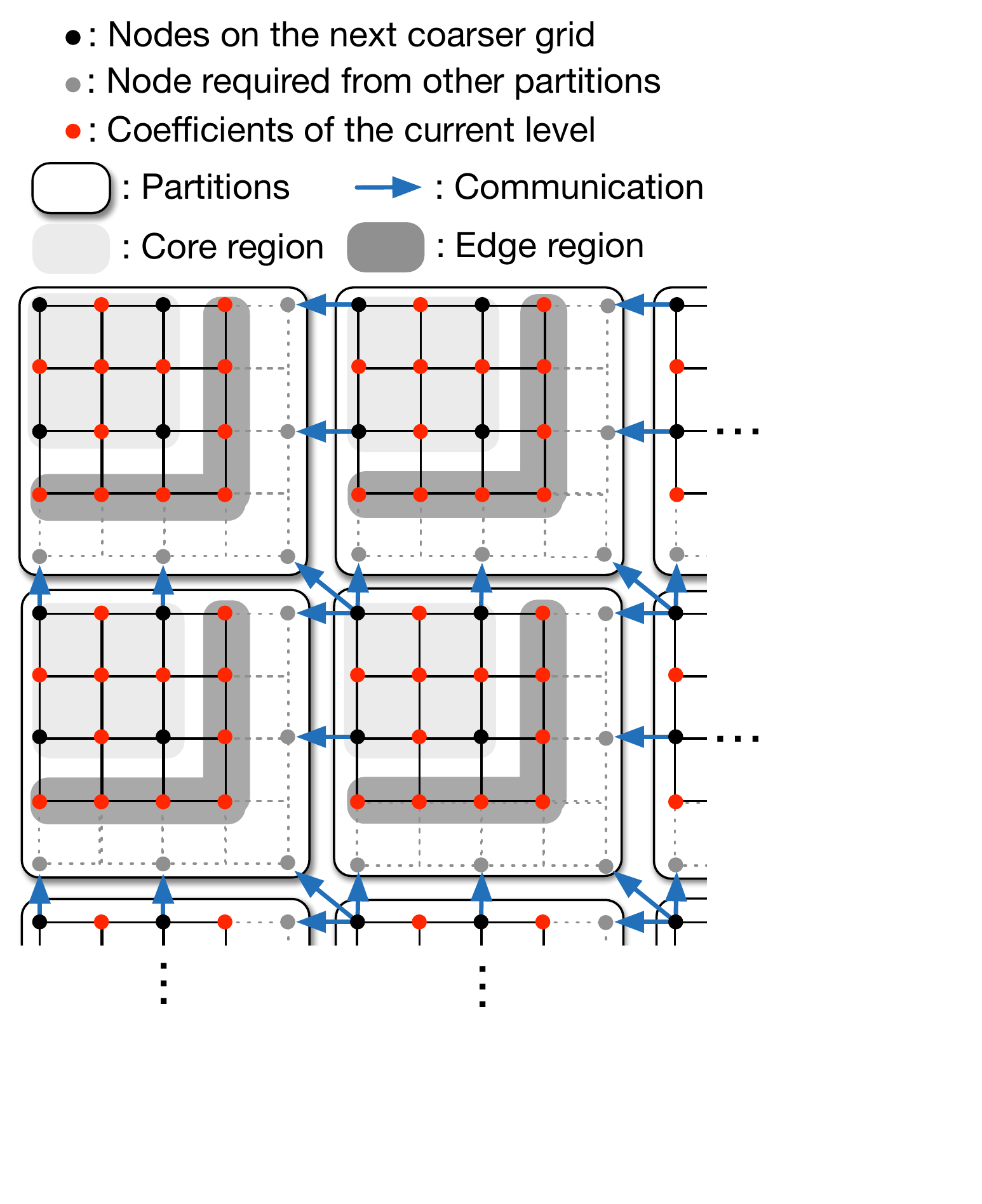}
    \caption{Cooperative parallel \gpk~and \lpk~ on multiple GPUs}
    \label{para-gpk-lpk}
    \vspace{-2em}
\end{figure}

\subsubsection{Cooperative parallel grid processing kernel}
\gpk{} is used for calculating the multilinear interpolations at nodes in $N_l \setminus N_{l-1}$ using nodal values in $N_l$. 
Essentially, each interpolation at node in $N_l \setminus N_{l-1}$ needs the nodal values of its neighboring nodes.
So, in our cooperative parallel design of \gpk{}, we add communications between GPUs to enable exchanging nodal values near the boundaries of each data partition for interpolation on nodes near boundaries.
Since communication only involves nodes on the boundary, the communication cost is low i.e., $O(n^{\frac{2}{3}})$ for 3D data.
Also, the interpolations on the nodes not close to the boundaries (core region) can be done without requiring exchanging nodal values, so we can overlap the interpolation on the core region with communications on the edge region.
\subsubsection{Cooperative parallel linear processing kernel}
\lpk{} shares similar data dependencies with \gpk{}, except it only needs to exchange boundary nodal values along dimensions it is currently processing.
We can also apply similar computation-communication overlapping.

\begin{figure}[h!]
    \vspace{-1em}
    \centering
    \includegraphics[width=0.3\textwidth, trim=0mm 0mm 0mm 0mm, clip]{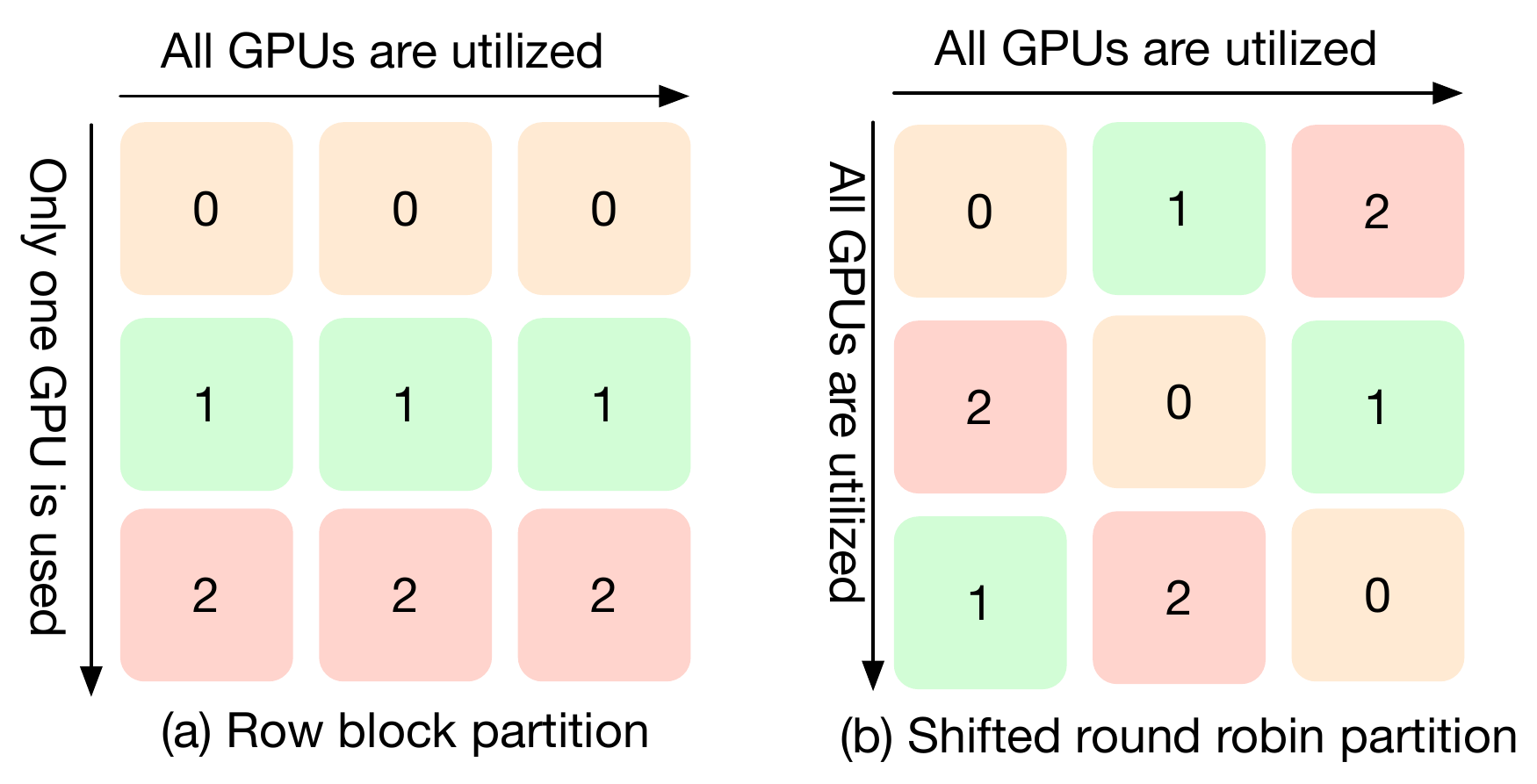}
    \caption{Illustration of different partitioning strategies used in cooperative parallel \ipk{} with 3 GPUs}
    \label{round-robin}
    \vspace{-2em}
\end{figure}

\subsubsection{Cooperative parallel iterative processing kernel}
Unlike \gpk{} and \lpk{}, strong data dependencies exist between different operations in \ipk.
In data refactoring \ipk{}is used for solving the corrections.
It requires a forward and backward pass along each dimension.
So each partition needs to be processed in an orderly fashion.
In this case, partitioning the data either row-block wise or column-block wise can lead to serious GPU underutilization.
For example, as shown in figure~\ref{round-robin}(a), when using \ipk{} on the second dimension, the work on the three GPUs needs to be done sequentially. 
To maximize GPU utilization, we propose to partition in a shifted round-robin fashion as shown in figure~\ref{round-robin}(b), so that all GPUs can be kept busy no matter which dimension \ipk{} is processing.

%% file: tex/evaluation.tex
\section{Experimental Evaluation}\label{sec:eval}
We evaluate our work on two GPU-enabled platforms.
Each node of the \textbf{Summit} supercomputer at ORNL is equipped with 6 NVIDIA \textbf{Volta} GV100 GPUs 
with 16 GB memory on each GPU and two 22-core (of which 21 cores/socket are accessible for computation) IBM POWER9 CPUs with 512 GB memory. 
\textbf{Turing} is a GPU-accelerated desktop with an NVIDIA RTX 2080 Ti GPU with 11 GB of memory and one 8-core Intel i7-9700K CPU with 32 GB of memory.

\begin{figure}[ht!]
    \vspace{-1em}
    \centering
    \begin{subfigure}[t]{0.5\textwidth}
    \includegraphics[width=\textwidth]{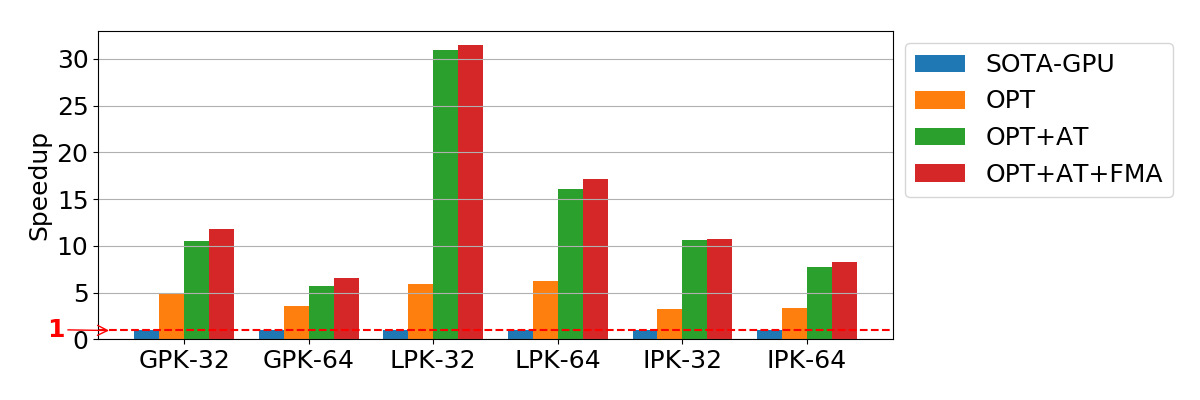}
    \vspace{-2em}
    \caption{Summit}
    \end{subfigure}
    \begin{subfigure}[t]{0.5\textwidth}
    \includegraphics[width=\textwidth]{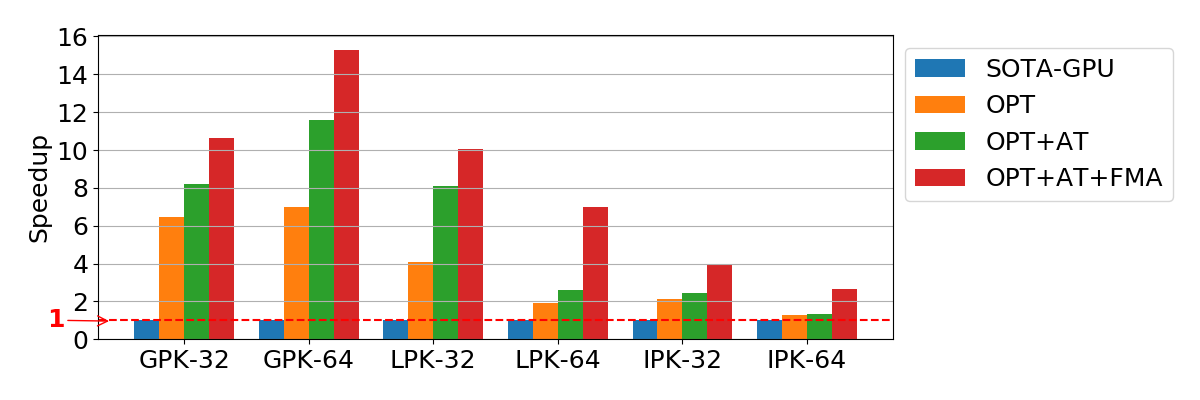}
    \vspace{-2em}
    \caption{Turing}
    \end{subfigure}
    \caption{Speedups achieved through using our proposed processing kernels compared with the state-of-the-art GPU designs. 32 and 64 represent single and double precision input.}
    \label{kernel-speedup}
    \vspace{-1em}
\end{figure}

\begin{figure}[ht]
    \centering
     \includegraphics[width=0.9\columnwidth]{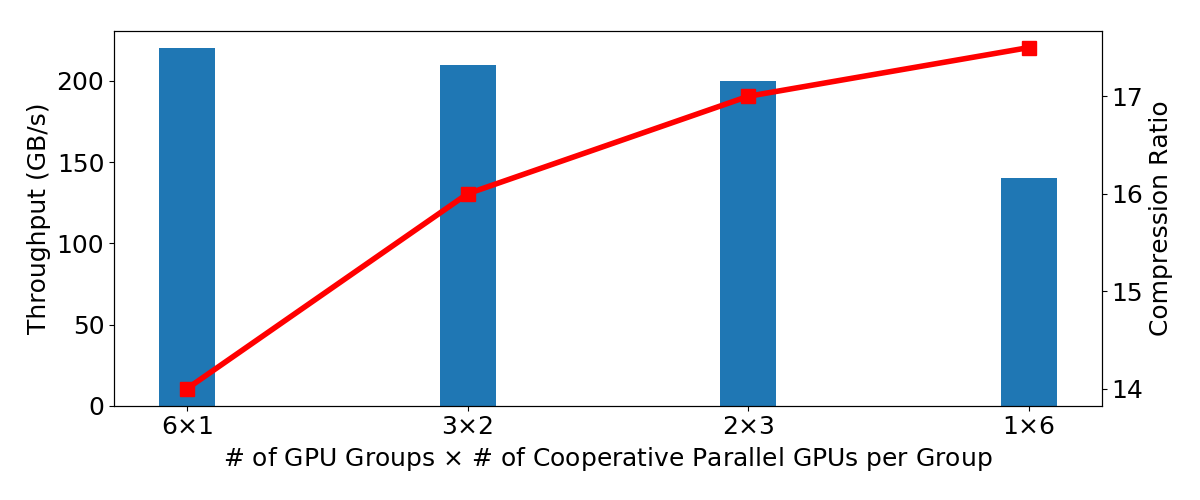}
     \vspace{-1em}
    \caption{Trade-off between compression throughput and ratio when using cooperative parallel GPU data refactoring. All 6 GPUs on a Summit node are used and grouped into several embarrassingly parallel GPU groups. GPUs in each group are using cooperative parallel approach.}
    \label{mgpu}
    
\end{figure}

\begin{figure}[ht]
    \centering
     \includegraphics[width=0.9\columnwidth]{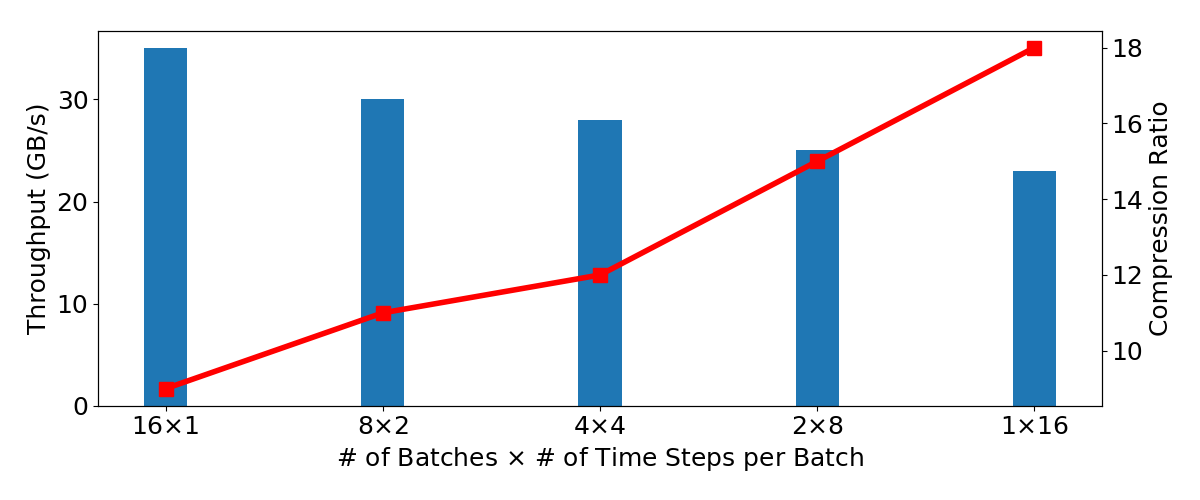}
     \vspace{-1em}
    \caption{Trade-off between compression throughput and ratio when use GPU data refactoring optimized for  spatiotemporal data. Total 16 time steps of data are divided into batches along the temporal dimension. }
    \label{mtime}
\end{figure}

\subsection{Evaluation methodology}
%\subsubsection{Input datasets}
We use datasets from a Gray-Scott reaction-diffusion simulation \cite{pearson1993complex, gscode}.
Each node in the input grid data is represented as single or double precision floating point values.
Note that our data refactoring algorithms have deterministic computation time complexity regardless of the values in the chosen dataset, so it will yield the same performance for any dataset with the same dimensions and size.
For simplicity, we let each dimension have the same size in our experiments.
% We configure the simulation codes to generate 3D data so that each dimension is in the form of $2^L+1$, where $L$ is an positive integer and is not necessary the same for all dimensions.
% If at least of one dimension is not in this form, one extra pre-processing step and the corresponding post-processing step are necessary, which consist of one iteration of special decomposition and recomposition.
% Since the extra one iteration for pre-processing and post-processing contributes a small portion of the total execution time, in order to better show the results of our optimizations on the main decomposition/recomposition loop we avoid those step in our tests by generating data with dimensions that follows the form of $2^L+1$.

%\subsubsection{Data refactoring implementations}
We evaluate five different data refactoring implementations.
\begin{itemize}
    \item \textbf{SOTA-GPU}: We use the state-of-the-art GPU data refactoring in the MGARD lossy compression software~\cite{mgard} as our GPU baseline. Its design includes two performance tuning parameters: thread block size and number of CUDA streams. In our evaluation, we use the best performance achieved by hand tuning those parameters.
    \item \textbf{SOTA-CPU}: We use the state-of-the-art CPU data refactoring implemented in the MGARD lossy compression software~\cite{mgard}, parallelized with MPI for a fair comparison, as our CPU baseline.
    \item \textbf{OPT}: Our GPU data refactoring, which uses our novel processing kernels (i.e., \gpk,~\lpk, and~\ipk) proposed in our previous work~\cite{chen2020accelerating}.
    \item \textbf{OPT+AT}: Our GPU data refactoring with auto tuning proposed in our previous work~\cite{chen2020accelerating}.
    \item \textbf{OPT+AT+FMA}: Our GPU data refactoring with auto tuning and FMA instruction optimizations.
    \item \textbf{OPT+AT+FMA+REO}: Our GPU data refactoring with auto tuning, FMA instruction optimizations, and our improved reordered data layout.
\end{itemize}

\subsection{Evaluation on kernels}

We first show the performance improvement we achieve from accelerating the three major operations in data refactoring on GPUs.
Figure~\ref{kernel-speedup} shows speedups achieved on the three operations on the two GPU platforms with both single and double precision inputs. 
The input size is 513$\times$513$\times$513.
For single precision input, with the thread-level load-compute decoupled design, coefficient calculation with \gpk{} outperforms the existing design by 4.9$\times$ and 6.9$\times$ on Summit and Turing GPUs, respectively.
For mass-transfer matrix multiplication, with higher thread concurrency and data dependency free calculation, \lpk{} achieves 6.3$\times$ and 4.1$\times$ speedups on Summit and Turing GPUs, respectively. 
For correction solver, \ipk{} triples the performance on Summit and doubles the performance on Turing with the same level of thread concurrency as the state-of-the-art design, thanks to the more efficient memory access patterns.
Also, leveraging our heuristic auto tuning capability, the optimum configurations can be selected automatically, yielding additional 1.2--4.9$\times$ speedups compared with choosing one configuration for all kernels and input sizes.

\subsection{Evaluation on optimization for consumer-class GPUs on edge systems}
As shown in Figure~\ref{kernel-speedup}, comparing with the speedups on Volta GPU on Summit, we observe relative lower speedups for our Turing GPU, since the consumer-class GPU has limited computational power especially for high precision data.
In this case, the computation can to be wrongly converted to compute bound.
Our FMA instruction optimization allows us to eliminate expensive operations, reducing warp latency and register usage, yielding additional 1.3-2.7$\times$ speedups on the three kernels.

% As our linear processing framework involves complicated designs, we study the mass matrix multiplication kernel designed following the framework to show how our optimization impacts performance. 
% Figure~\ref{linear-opt}} shows the memory throughput of mass matrix multiplication as different optimizations are applied.
% Specifically, we show the performance of mass matrix multiplication kernel when decomposing a 4097$\times$4097 grid of input data, which needs 12 levels of decomposition.
% We can see the original serial CPU version suffers from degraded performance due to inefficient memory access when grid spacing is large (level is small).
% This is also the case for the naive GPU design.
% \added{The naive GPU design parallelizes the workload vector-wise \cite{Basu:2017} without applying memory access efficiency optimizations proposed in this work.}
% When mass multiplication kernel are designed following our linear processing framework we can see it can achieve much better performance and can sustain similar performance when grid spacing is large (i.e., $l$ is small) and only degrades on small grids, which bring minor impacts to the overall performance.

% \begin{figure}[ht!]
%     \centering
%     \includegraphics[width=0.43\textwidth]{figures/linear-smt.pdf}
%     \caption{Using mass matrix multiplication to show how much performance can be improved by designing kernels following our linear processing framework (evaluated on a single NVIDIA Tesla V100 GPU on Summit)}
%     \label{linear-opt}
%     \vspace{-1em}
% \end{figure}

\subsection{Evaluation on overall data refactoring on a single GPU}
\begin{figure*}[ht]
    \centering
    \begin{subfigure}[t]{0.45\textwidth}
    \includegraphics[width=\textwidth]{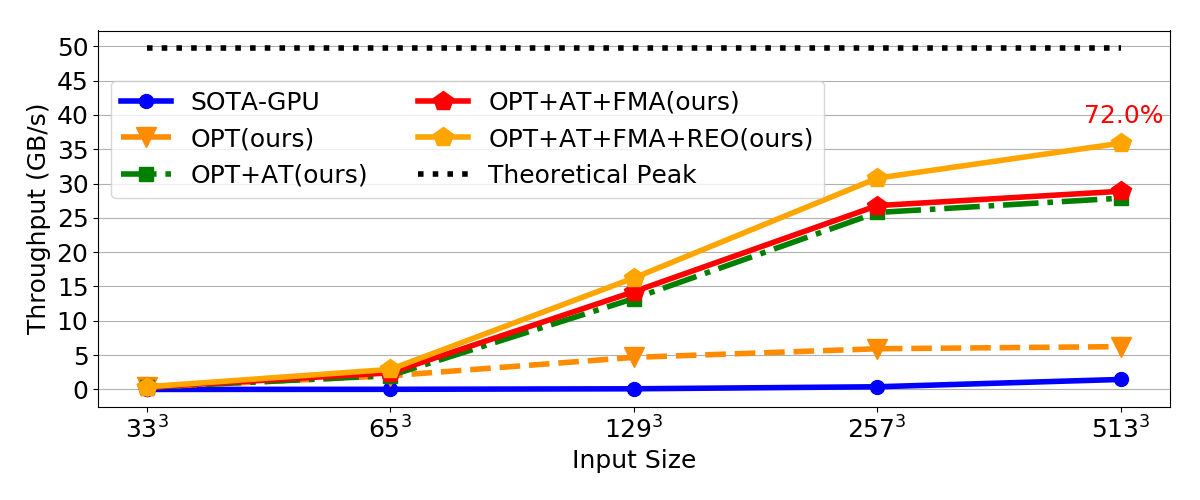}
    \vspace{-2em}
    \caption{Volta (32-bit)}
    \end{subfigure}
    \begin{subfigure}[t]{0.45\textwidth}
    \includegraphics[width=\textwidth]{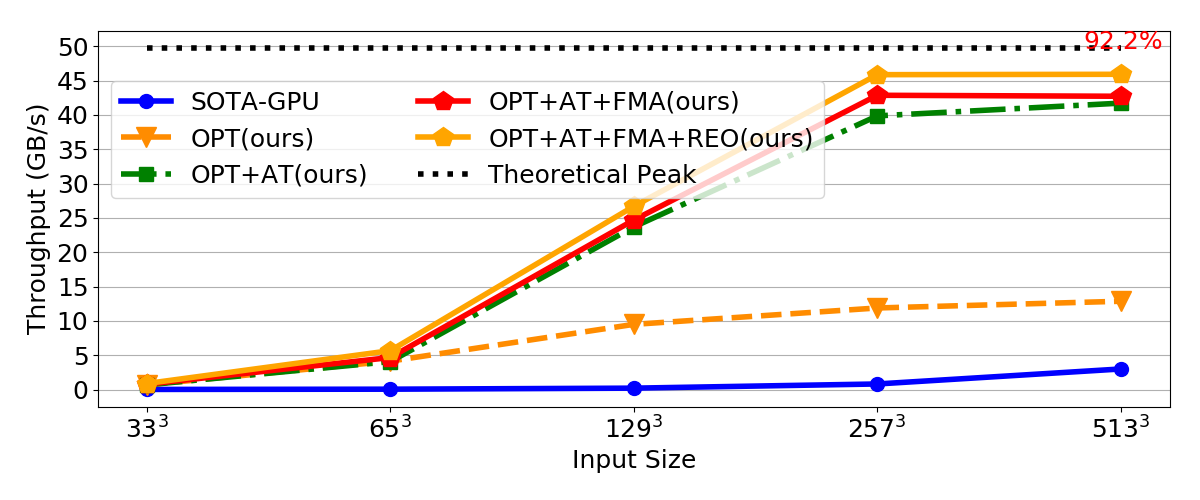}
    \vspace{-2em}
    \caption{Volta (64-bit)}
    \end{subfigure}
    \begin{subfigure}[t]{0.45\textwidth}
    \includegraphics[width=\textwidth]{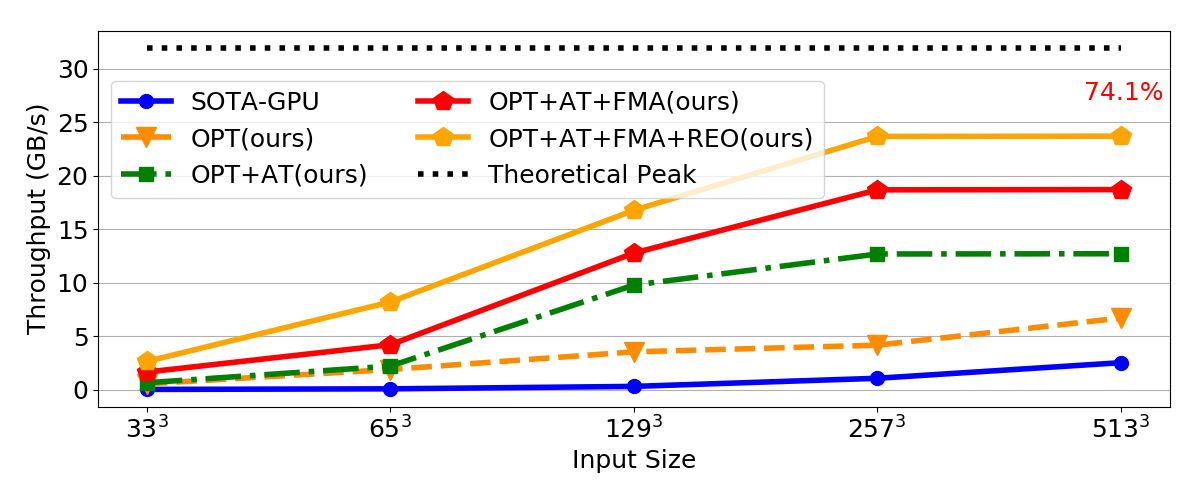}
    \vspace{-2em}
    \caption{Turing (32-bit)}
    \end{subfigure}
    \begin{subfigure}[t]{0.45\textwidth}
    \includegraphics[width=\textwidth]{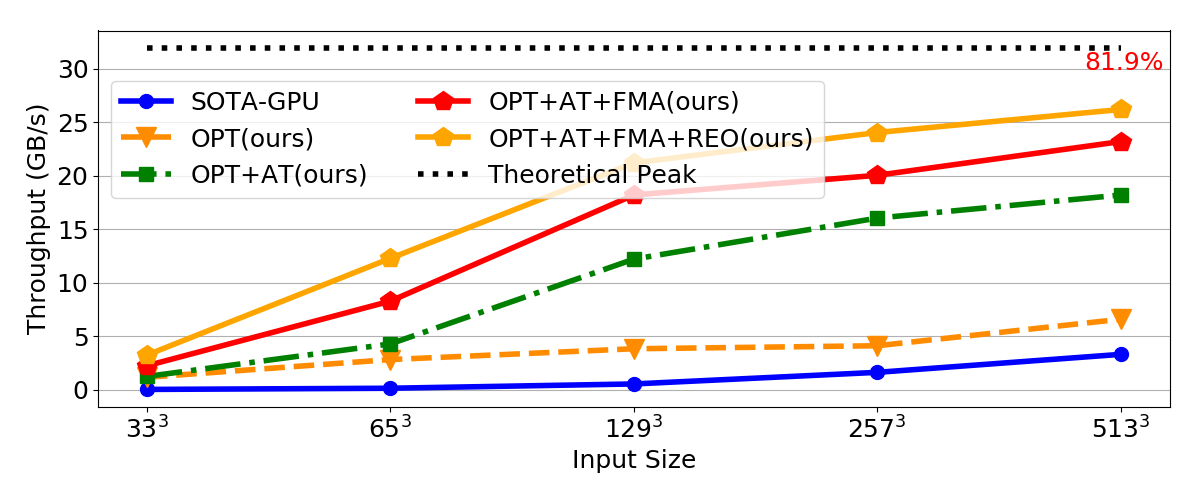}
    \vspace{-2em}
    \caption{Turing (64-bit)}
    \end{subfigure}
    \caption{Data refactoring throughput on a single GPU}
    \label{fig:overall2}
    \vspace{-2em}
\end{figure*}

Figure~\ref{fig:overall2} shows the end-to-end data refactoring throughput achieved on a single GPU with different input sizes.
(As decomposition and recomposition are symmetric process, they have identical performance.)
To see how close the achieved data refactoring throughput is to the theoretical peak throughput, we estimate the theoretical peak by dividing the achievable single pass throughput
%\footnote{The achievable single pass throughput is the maximum throughput achievable when passing (reading and storing) data on GPU memory for one time. We measured it through a specially design benchmark kernel that simultaneously reads and writes the same amount of data from and to the GPU memory without computation.}
%\ian{I think footnotes are ugly.} 
with the accumulated number of passes on the entire input data over the data refactoring process. 
(The achievable single pass throughput is the maximum throughput achievable when data are read and stored on GPU memory once. We measured it through a specially designed benchmark kernel that simultaneously reads and writes the same amount of data from and to the GPU memory without computation.)
%The bandwidth is calculated by dividing the size of the data by kernel time. Specifically, we achieved $\sim$260 GB/s and $\sim$420 GB/s for reading/writing throughput on Turing and Volta GPU.} by the accumulated number of passes on input data over the entire data refactoring process:
The accumulated number of passes is calculated by summing the number of passes for all decomposition levels: \textit{passes per level} $\times \frac{1}{1-\frac{1}{8}}$. \textit{passes per level} = 1(coefficient calculation) + 1(copy to workspace) + 5.25(correction calculation) + 0.125(apply correction). 
The theoretical peaks for Summit and Turing GPUs
are 49.8 GB/s and 32.0 GB/s, respectively, for both single and double precision data. 
The state-of-the-art GPU data refactoring methods that we use as our baseline achieve only up to 10.4\% of the theoretical peak throughput;
our optimized GPU data refactoring achieves up to 92.2\% of theoretical peak.

\subsection{Evaluation on single-node multi-GPU performance}
To show the potential of taking advantage of dense multi-GPU architecture using our cooperative parallel data refactoring on GPUs, we conduct an evaluation using 6 GPUs on one Summit node.
We integrate our GPU data refactoring the MGARD compression workflow to demonstrate the benefit of using cooperative parallel data refactoring in terms of compression ratio.
We use a 16 GB dataset from the Gary-Scott simulation as input with $1e^{-3}$ error bound.
From the architecture level, the 6 GPUs are grouped into two GPU islands with 3 GPUs in each island interconnected using fast NVLink. 
Inter-island communication is achieved using X-Bus.
In our evaluation, we logically group the 6 GPUs into $K$ GPUs groups and do cooperative parallel data refactoring among the $S$ GPUs within each group and keep embarrassingly parallel across GPU groups. 
We evaluate different combinations and note them as $K\times S$.
As shown in figure~\ref{mgpu}, in terms of performance, we can see $6\times 1$ offers the maximum throughput since all 6 GPUs are refactoring data in an embarrassingly parallel fashion.
$3 \times 2$ has slightly lower throughput compared with $6\times 1$ since every two GPUs need to communicate with each other.
We see similar throughput between $3 \times 2$ and $2 \times 3$.
When all 6 GPUs are participating in cooperative parallel ($1\times 6$) we observe noticeable performance degradation due to the heavy communication across the X-Bus.
In terms of compression ratios, as we increase the number of GPUs participating in cooperative parallel data refactoring, we observe improvement in compression ratios since compression can take more advantage of the data correlation between partitions. 

\subsection{Evaluation on optimization for spatiotemporal data refactoring}
Next, we evaluate our optimization for refactoring spatiotemporal data on GPUs.
For input, we use 16 time steps of Gray-Scott simulation data with 512 MB data per time step.
We use one GPU for refactoring and integrate data refactoring the MGARD compression workflow.
The compression error bound is set to $1e^{-3}$.
We demonstrate the trade-off between compression throughput and ratio when batching different numbers of time steps of simulation data for compression each time.
As we can see in figure~\ref{mtime}, when we increase the number of time step per compression batch, we effectively improve the compression ratio.
However, we also observe performance degradation as we increase the batch size.
This is anticipated as exploiting the temporal dimension leads to extra data refactoring costs.

\subsection{Evaluation on multi-node multi-GPU performance}
To show the potential of GPU-accelerated data refactoring in large-scale scientific applications, we conduct a weak scaling test on Summit. 
We parallelize the workload by assigning each GPU or CPU core an equal-sized data partition and perform decomposition and recomposition.

\begin{figure}[ht]
    \vspace{-1em}
    \centering
     \includegraphics[width=0.9\columnwidth]{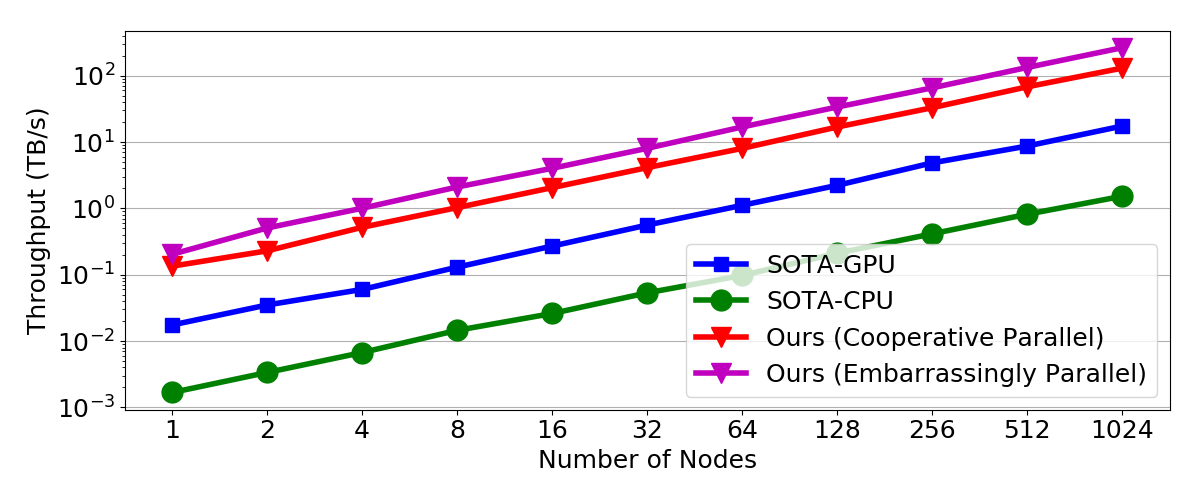}
     \vspace{-1em}
    \caption{Aggregated data refactoring throughput at scale on Summit. 6 GPUs or 42 CPU cores are used per computing node, with each GPU or CPU core handling 1 GB in double precision.}
    \label{scale-smt}
    \vspace{-2em}
\end{figure}

%\dave{Calling the performance "great" reads funny. This should be quantified, if possible.}
We assign each GPU device or CPU core to one MPI process and perform data refactoring on 1 GB of simulation data.
For each computing node, we use the total available number of GPUs and CPU cores.
We scale the number of nodes up to 1024 in our tests on Summit.
For both SOTA-CPU and SOTA-GPU, we use embarrassingly parallel approach and we test both embarrassingly parallel and cooperative parallel approach for our work.
Cooperative parallel is restricted within each computing node, since inter-node communication can bring high overhead.
As shown in Figure~\ref{scale-smt}, our optimized GPU data refactoring method achieves much greater throughput than state-of-the-art GPU and CPU designs. 
For example, we need only four computing nodes to achieve 1 TB/s refactoring throughput, 
whereas state-of-the-art GPU and CPU designs require 64 and 512 nodes, respectively. 
With 1024 nodes (i.e., 6144 Volta GPUs), we achieve up to 130 TB/s and 264 TB/s aggregated data refactoring throughput with cooperative parallel and embarrassingly parallel approach, respectively. 
%These numbers show great potential in speeding up data refactoring-based I/O operations.
%especially when used in combination with emerging new technologies that help move data in or out of the GPU efficiently, such as GPU-CPU NVLink and GPUDirect Storage.

%% file: tex/showcase.tex
\begin{figure*}[ht]
    \centering
    \begin{subfigure}[t]{0.45\textwidth}
    \includegraphics[width=\textwidth]{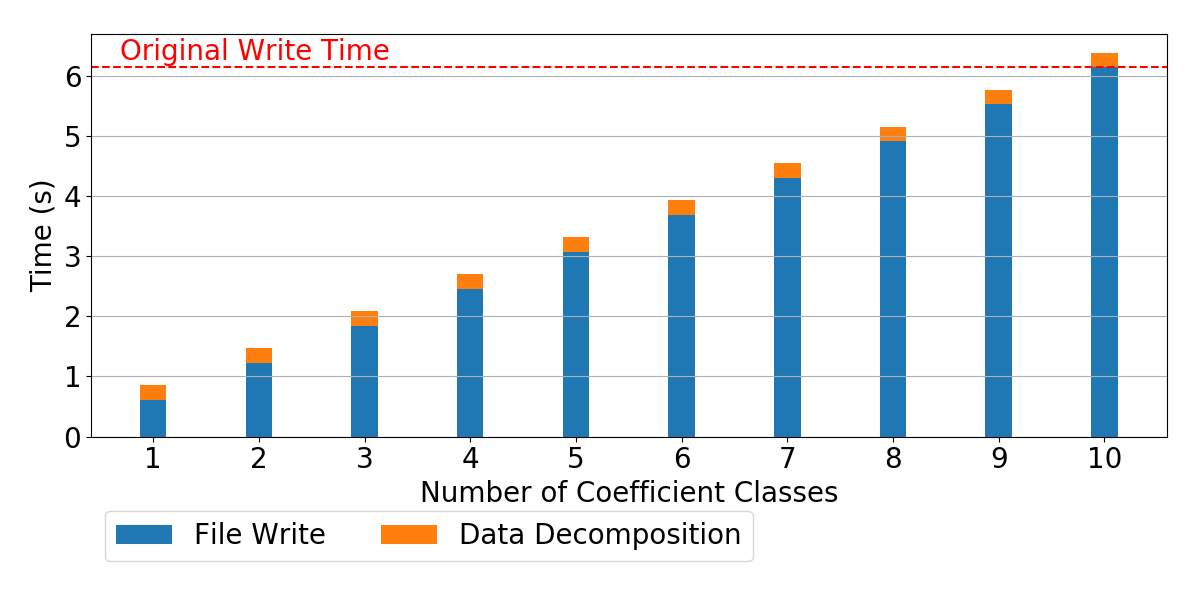}
    \vspace{-2em}
    \caption{Write simulation data}
    \end{subfigure}
    \begin{subfigure}[t]{0.45\textwidth}
    \includegraphics[width=\textwidth]{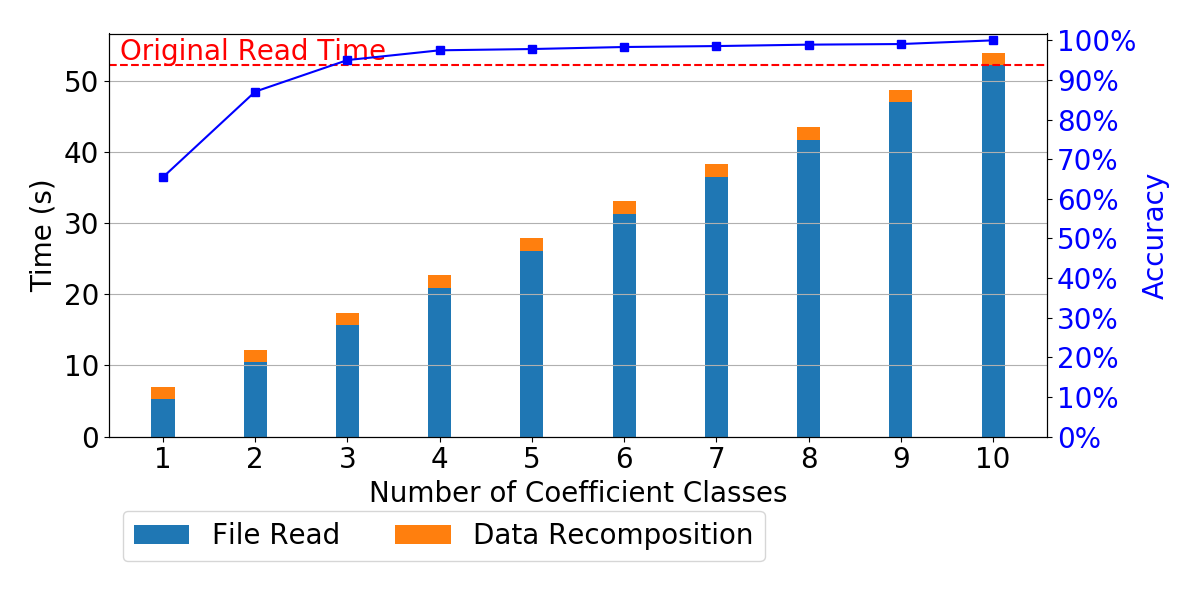}
    \vspace{-2em}
    \caption{Read simulation data and visualize}
    \end{subfigure}
    \caption{Showcase 1: Data refactoring in scientific visualization workflow}
    \label{vis-showcase}
    \vspace{-1em}
\end{figure*}

\begin{figure*}[ht]
    \centering
    \begin{subfigure}[t]{0.45\textwidth}
    \includegraphics[width=\textwidth]{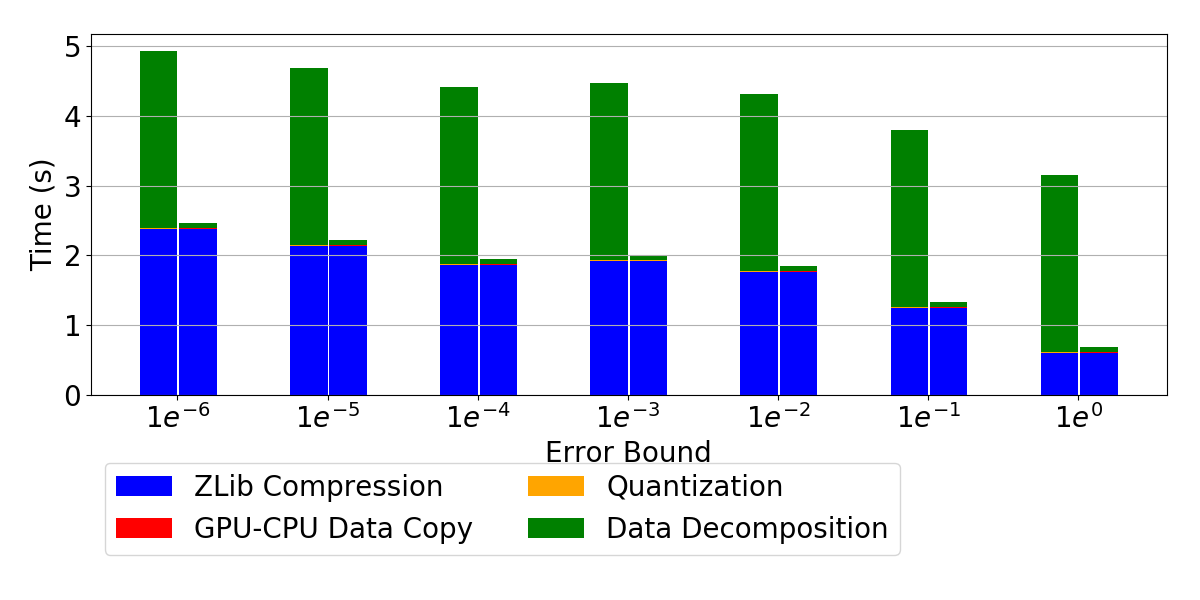}
    \vspace{-2em}
    \caption{Compression}
    \end{subfigure}
    \begin{subfigure}[t]{0.45\textwidth}
    \includegraphics[width=\textwidth]{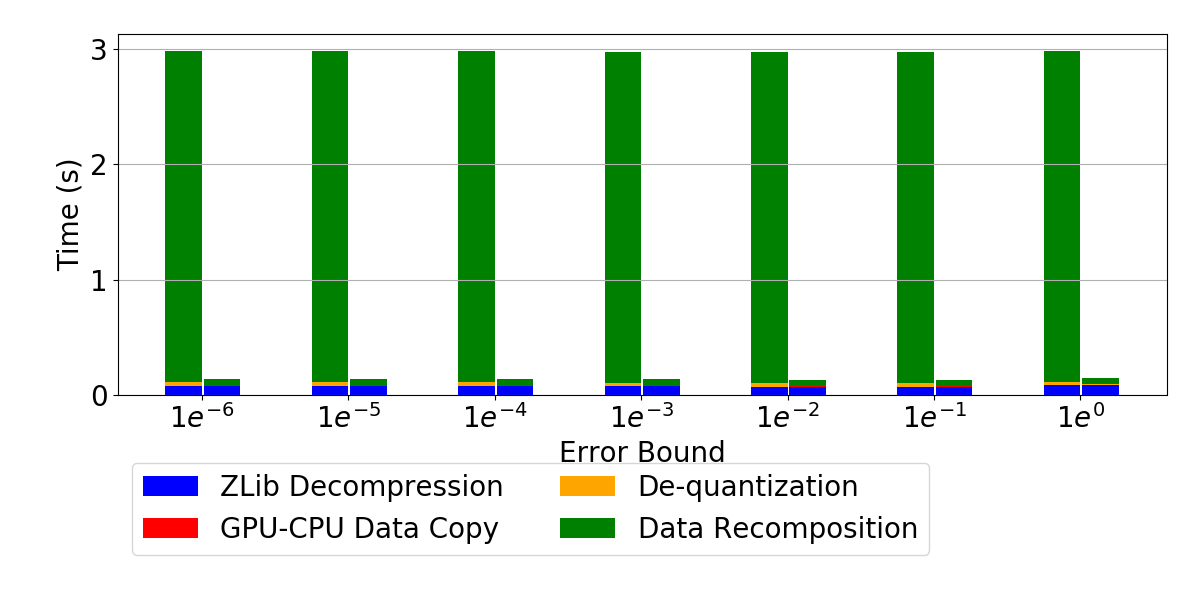}
    \vspace{-2em}
    \caption{Decompression}
    \end{subfigure}
    \caption{Showcase 2: MGARD lossy compression using CPU (left bars) vs.\ GPU (right bars)}
    \label{mgard-showcase}
    \vspace{-1.5em}
\end{figure*}

\section{Showcase}\label{sec:showcase}
Data refactoring algorithms were designed to offer much greater flexibility when managing large scientific data than the traditional methods.
With well-designed data management, data can be shared between scientific applications more intelligently with a large reduction in I/O costs.
However, inefficient data refactoring routines can diminish the benefits brought by data refactoring itself. 
Here we use two examples to show the benefits of GPU-based data refactoring over the CPU designs.

\subsection{Visualization workflow}
First we show how our GPU optimizations can make data refactoring effective when used for I/O cost reduction in scientific workflows that rely on file-based data sharing.
Figure~\ref{vis-showcase} shows the cost of writing and reading a 4 TB simulation data file using 4096 and 512 processes using the state-of-the-art ADIOS I/O library \cite{liu2014hello} on Summit with GPU-accelerated data refactoring enabled.
By writing or reading fewer coefficient classes, we can see immediate cost reduction in file write and read.
When our efficient GPU-accelerated data refactoring is used, we can see this reduction in the cost of file write and read can be effectively translated into a reduction in the total I/O cost.
Although multigrid-based data refactoring allows us to encode the most important information in the data with a few coefficient classes, it would not reduce the total I/O cost unless those coefficient classes can be efficiently computed or used for data recovery.
For example, in our experiments we achieve $\sim$95$\%$ accuracy for a chosen feature in the visualization result (i.e., the total area of the iso-surfaces~\cite{chen2019understanding, yakushin2020feature}) with only three out of ten coefficient classes.
This can be effectively translated into $\sim$66$\%$ I/O cost reduction.

\subsection{Lossy compression}
Multigrid-based hierarchical data refactoring can also be used as a preconditioner in scientific lossy compression software.
As one of the key components in lossy compression workflows, it is important to have efficient data refactoring in order to make fast lossy compression possible.
We showcase how our GPU-accelerated data refactoring can help improve the performance of lossy compression workflows in the MGARD lossy compression software.
MGARD is a CPU-based lossy compressor with three components in its workflow: multigrid-based data refactoring, quantization, and entropy encoding.
Figure~\ref{mgard-showcase} shows the time breakdown of the each component in MGARD \cite{mgard} when data refactoring remains on the CPU (left bars) or is off-loaded to the GPU (right bars).
In our test, besides the data refactoring process, we also off-load the quantization and de-quantization processes to the GPUs, since it can help reduce the GPU-CPU data transfer cost.
The entropy encoding stage (ZLib lossless compression) is kept on the CPU.
We can see that our GPU-accelerated data refactoring can greatly reduce the overall execution time of the lossy compression workflows.

%% file: tex/related_works.tex
\section{Related Work}

Multigrid-based data refactoring shares some similarities with multigrid solvers,
such as the use of multiple interlocking grids.
%However, the different goals of data refactoring and solvers makes it hard to apply
%GPU optimizations designed for the former to the latter.
%This is due to the fact that data refactoring is designed and optimized to target a different goal
But while multigrid solvers aim to accelerate the solving of linear systems,
multigrid-based data refactoring aims to reconstruct scientific data progressively with hierarchical representations. 
This difference in focus leads to fundamental differences in both algorithms and optimization that prevent direct translation of
GPU optimizations. % designed for the former to the latter..

From an algorithmic perspective,
although data refactoring and multigrid solvers have some operations in common, data refactoring composes these operations in a unique way. 
Further, the correction used in data refactoring is designed specifically for the orthogonal projection, 
while the correction in multigrid solvers is used to generate the fine grid solution. From GPU optimization perspective: optimizations for data refactoring need to consider handling large-volume scientific data, which means we need to consider not only limited GPU memory but also cases where refactoring process might share resources with original scientific computations on GPUs. So, it is essential to optimize for low memory footprint as well as performance.
Although part of the kernels used in data refactoring share similar computation patterns to those found in multigrid solvers, it is challenging to leverage existing work directly to achieve good parallelism and memory footprint balance in data refactoring.
For example, state-of-the-art GPU refactoring~\cite{mgard} uses a parallelization technique proposed by Basu et al.~\cite{Basu:2017}, which only use coarse grain vector-wise parallelism, which can cause lower performance for data refactoring. 
Although fine-grain parallelism has been achieved in previous works~\cite{Bell:2012,Esler:2012,Sebastian:2014,Richter:2015,Clark:2016}, they generally brings high memory footprint and would require considerable efforts to apply their optimizations to different algorithms.

%% file: tex/conclusion.tex
\section{Conclusion}

%As I/O becomes a major performance bottleneck for scientific computing, data refactoring shows great potential to reduce I/O costs.
We have presented optimized data refactoring kernels that allow for use of GPUs to
accelerate multigrid-based hierarchical refactoring for scientific data.
%using highly optimized data refactoring kernels. % specialized for scientific data refactoring.
We evaluated our designs on two platforms, including the leadership-class Summit supercomputer at ORNL,
and showed that 
our GPU version can speed up data refactoring by up to 160$\times$ and 15$\times$ compared with state-of-the-art CPU and GPU designs, respectively,
and can achieve 264 TB/s throughput using 1024 nodes on Summit
We also showcased our work using a large-scale scientific visualization workflow and the MGARD lossy compression technique.  Together, these results demonstrate that scientists have another opportunity for dealing with their high data throughput requirements.  Inline refactoring of scientific data can offer performance improvements and temporal fidelity that can benefit a number of science scenarios.

% \section*{Acknowledgments}
%   Omitted for double-blind review.

%% This research was supported by the Department of Energy's SciDAC RAPIDS Institute (??????), as well as the Exascale Computing Project (17-SC-20-SC), a collaborative effort of U.S. Department of Energy Office of Science and the National Nuclear Security Administration. This research used resources of the Argonne and Oak Ridge Leadership Computing Facilities, DOE Office of Science User Facilities supported under Contracts DE-AC02-06CH11357 and DE-AC05-00OR22725, respectively.

%% file: main.bbl
\begin{thebibliography}{10}
\providecommand{\url}[1]{#1}
\csname url@samestyle\endcsname
\providecommand{\newblock}{\relax}
\providecommand{\bibinfo}[2]{#2}
\providecommand{\BIBentrySTDinterwordspacing}{\spaceskip=0pt\relax}
\providecommand{\BIBentryALTinterwordstretchfactor}{4}
\providecommand{\BIBentryALTinterwordspacing}{\spaceskip=\fontdimen2\font plus
\BIBentryALTinterwordstretchfactor\fontdimen3\font minus
  \fontdimen4\font\relax}
\providecommand{\BIBforeignlanguage}[2]{{%
\expandafter\ifx\csname l@#1\endcsname\relax
\typeout{** WARNING: IEEEtran.bst: No hyphenation pattern has been}%
\typeout{** loaded for the language `#1'. Using the pattern for}%
\typeout{** the default language instead.}%
\else
\language=\csname l@#1\endcsname
\fi
#2}}
\providecommand{\BIBdecl}{\relax}
\BIBdecl

\bibitem{alexander2020exascale}
F.~Alexander \emph{et~al.}, ``Exascale applications: Skin in the game,''
  \emph{Philosophical Transactions of the Royal Society A}, 2020.

\bibitem{Wan:2019}
L.~Wan \emph{et~al.}, ``Data management challenges of exascale scientific
  simulations: A case study with the {Gyrokinetic Toroidal Code} and {ADIOS},''
  in \emph{The 10th International Conference on Computational Methods}, 2019.

\bibitem{ku2009full}
S.~Ku \emph{et~al.}, ``Full-f gyrokinetic particle simulation of centrally
  heated global {ITG} turbulence from magnetic axis to edge pedestal top in a
  realistic tokamak geometry,'' \emph{Nuclear Fusion}, 2009.

\bibitem{chang2004numerical}
C.-S. Chang \emph{et~al.}, ``Numerical study of neoclassical plasma pedestal in
  a tokamak geometry,'' \emph{Physics of Plasmas}, 2004.

\bibitem{taylor2004science}
R.~Taylor \emph{et~al.}, ``Science with the {Square Kilometer Array}:
  Motivation, key science projects, standards and assumptions,'' \emph{arXiv
  preprint astro-ph/0409274}, 2004.

\bibitem{hpss}
``{HPSS},'' \url{www.hpss-collaboration.org/}. Accessed: 10/2020.

\bibitem{ainsworth2018multilevel}
M.~Ainsworth \emph{et~al.}, ``Multilevel techniques for compression and
  reduction of scientific data—the univariate case,'' \emph{Computing and
  Visualization in Science}, 2018.

\bibitem{ainsworth2019multilevel}
M.~Ainsworth \emph{et~al.}, ``Multilevel techniques for compression and
  reduction of scientific data---the multivariate case,'' \emph{SIAM J.\
  Scientific Computing}, 2019.

\bibitem{ainsworth2019multilevel2}
M.~Ainsworth \emph{et~al.}, ``Multilevel techniques for compression and
  reduction of scientific data---quantitative control of accuracy in derived
  quantities,'' \emph{SIAM J.\ Scientific Computing}, 2019.

\bibitem{CODAR2020}
I.~Foster \emph{et~al.}, ``Online data analysis and reduction: An important
  co-design motif for extreme-scale computers,'' \emph{International Journal of
  High-Performance Computing Applications}, 2020.

\bibitem{li2018tartan}
A.~Li \emph{et~al.}, ``Tartan: evaluating modern gpu interconnect via a
  multi-gpu benchmark suite,'' in \emph{2018 IEEE International Symposium on
  Workload Characterization (IISWC)}.\hskip 1em plus 0.5em minus 0.4em\relax
  IEEE, 2018.

\bibitem{li2019evaluating}
A.~Li \emph{et~al.}, ``Evaluating modern gpu interconnect: Pcie, nvlink,
  nv-sli, nvswitch and gpudirect,'' \emph{IEEE Transactions on Parallel and
  Distributed Systems}, 2019.

\bibitem{mass}
``Mass matrix computation in the finite element method,''
  \url{https://demonstrations.wolfram.com/MassMatrixComputationInTheFiniteElementMethod}.

\bibitem{liang2020optimizing}
X.~Liang \emph{et~al.}, ``Optimizing multi-grid based reduction for efficient
  scientific data management,'' \emph{arXiv preprint arXiv:2010.05872}, 2020.

\bibitem{mgard}
\BIBentryALTinterwordspacing
\emph{{MGARD} Lossy Compression Software}, 2020 (accessed April 21, 2020).
  [Online]. Available: \url{https://github.com/CODARcode/MGARD}
\BIBentrySTDinterwordspacing

\bibitem{atkinson1985elementary}
K.~E. Atkinson \emph{et~al.}, \emph{Elementary numerical analysis}.\hskip 1em
  plus 0.5em minus 0.4em\relax Wiley New York, 1985.

\bibitem{chen2019tsm2}
J.~Chen \emph{et~al.}, ``{TSM2}: optimizing tall-and-skinny matrix-matrix
  multiplication on {GPU}s,'' in \emph{ACM International Conference on
  Supercomputing}, 2019.

\bibitem{rivera2021tsm2x}
C.~Rivera \emph{et~al.}, ``Tsm2x: High-performance tall-and-skinny
  matrix--matrix multiplication on gpus,'' \emph{Journal of Parallel and
  Distributed Computing}, vol. 151, pp. 70--85, 2021.

\bibitem{chen2016online}
J.~Chen \emph{et~al.}, ``Online algorithm-based fault tolerance for cholesky
  decomposition on heterogeneous systems with gpus,'' in \emph{2016 IEEE
  International Parallel and Distributed Processing Symposium (IPDPS)}.\hskip
  1em plus 0.5em minus 0.4em\relax IEEE, 2016, pp. 993--1002.

\bibitem{chen2018fault}
J.~Chen \emph{et~al.}, ``Fault tolerant one-sided matrix decompositions on
  heterogeneous systems with gpus,'' in \emph{SC18: International Conference
  for High Performance Computing, Networking, Storage and Analysis}.\hskip 1em
  plus 0.5em minus 0.4em\relax IEEE, 2018, pp. 854--865.

\bibitem{chen2016gpu}
J.~Chen \emph{et~al.}, ``Gpu-abft: Optimizing algorithm-based fault tolerance
  for heterogeneous systems with gpus,'' in \emph{2016 IEEE International
  Conference on Networking, Architecture and Storage (NAS)}.\hskip 1em plus
  0.5em minus 0.4em\relax IEEE, 2016, pp. 1--2.

\bibitem{chen2020accelerating}
J.~Chen \emph{et~al.}, ``Accelerating multigrid-based hierarchical scientific
  data refactoring on gpus,'' \emph{arXiv preprint arXiv:2007.04457}, 2020.

\bibitem{pearson1993complex}
J.~E. Pearson, ``Complex patterns in a simple system,'' \emph{Science}, 1993.

\bibitem{gscode}
``{Gray-Scott Simulation Code},'' \url{https://github.com/pnorbert/adiosvm/tree
  /master/Tutorial/gray-scott}.

\bibitem{liu2014hello}
Q.~Liu \emph{et~al.}, ``Hello {ADIOS}: The challenges and lessons of developing
  leadership class {I/O} frameworks,'' \emph{Concurrency and Computation:
  Practice and Experience}, 2014.

\bibitem{chen2019understanding}
J.~Chen \emph{et~al.}, ``Understanding performance-quality trade-offs in
  scientific visualization workflows with lossy compression,'' in \emph{2019
  IEEE/ACM 5th International Workshop on Data Analysis and Reduction for Big
  Scientific Data (DRBSD-5)}.\hskip 1em plus 0.5em minus 0.4em\relax IEEE,
  2019, pp. 1--7.

\bibitem{yakushin2020feature}
I.~Yakushin \emph{et~al.}, ``Feature-preserving lossy compression for in situ
  data analysis,'' in \emph{49th International Conference on Parallel
  Processing-ICPP: Workshops}, 2020, pp. 1--9.

\bibitem{Basu:2017}
P.~Basu \emph{et~al.}, ``Compiler-based code generation and autotuning for
  geometric multigrid on {GPU}-accelerated supercomputers,'' \emph{Parallel
  Computing}, 2017.

\bibitem{Bell:2012}
N.~Bell \emph{et~al.}, ``Exposing fine-grained parallelism in algebraic
  multigrid methods,'' \emph{SIAM J.\ Scientific Computing}, 2012.

\bibitem{Esler:2012}
K.~Esler \emph{et~al.}, ``{GAMPACK (GPU} accelerated algebraic multigrid
  package),'' in \emph{13th European Conference on the Mathematics of Oil
  Recovery}, 2012.

\bibitem{Sebastian:2014}
J.~Sebastian \emph{et~al.}, ``{GPU} accelerated three dimensional unstructured
  geometric multigrid solver,'' in \emph{International Conference on High
  Performance Computing and Simulation}, 2014.

\bibitem{Richter:2015}
C.~Richter \emph{et~al.}, ``Multi-{GPU} acceleration of algebraic multi-grid
  preconditioners for elliptic field problems,'' \emph{IEEE Transactions on
  Magnetics}, 2015.

\bibitem{Clark:2016}
M.~A. Clark \emph{et~al.}, ``Accelerating lattice {QCD multigrid on GPUs} using
  fine-grained parallelization,'' in \emph{SC'16}, 2016.

\end{thebibliography}
